\documentclass[a4paper,fleqn,usenatbib,useAMS]{mnras}

\usepackage{color}

\usepackage{graphicx}	
\usepackage{amsmath}	
\usepackage{amssymb}	
\usepackage{multicol}        
\usepackage{bm}		
\usepackage{pdflscape}	
\usepackage{adjustbox}
\usepackage{threeparttable}

\newcommand{\mean}[1]{\overline{#1}}
\defcitealias{goldreich_sridhar_1995}{GS95}
\defcitealias{lazarian_vishniac_1999}{LV99}
\defcitealias{karak_etal_2014}{K14}

\usepackage[T1]{fontenc}
\usepackage{ae,aecompl}
\usepackage{ulem}

\title[Diffusion of large-scale magnetic fields by reconnection in MHD turbulence]{Diffusion of large-scale magnetic fields by reconnection in MHD turbulence}

\author[Santos-Lima, Guerrero, de Gouveia Dal Pino \& Lazarian]{
	R. Santos-Lima,$^{1}$\thanks{Contact e-mail: \href{mailto:reinaldo.lima@iag.usp.br}{reinaldo.lima@iag.usp.br}}
	G. Guerrero,$^{2}$\thanks{Contact e-mail: \href{mailto:guerrero@fisica.ufmg.br}{guerrero@fisica.ufmg.br}}
	E. M. de Gouveia Dal Pino,$^{1}$\thanks{Contact e-mail: \href{mailto:dalpino@iag.usp.br}{dalpino@iag.usp.br}}
	A. Lazarian$^{3,4}$\thanks{Contact e-mail: \href{mailto:lazarian@astro.wisc.edu}{lazarian@astro.wisc.edu}}\\
$^{1}$Instituto de Astronomia, Geof\'isica e Ci\^encias Atmosf\'ericas, Universidade de S\~ao Paulo, R. do Mat\~ao, 1226, S\~ao Paulo, SP 05508-090, Brazil \\
$^{2}$Physics Department, Universidade Federal de Minas Gerais, Av. Antonio Carlos, 6627, Belo Horizonte, MG, Brazil, 31270-901 \\
$^{3}$Department of Astronomy, University of Wisconsin, 475 North Charter Street, Madison, WI 53706, USA \\
$^{4}$Center for Computation Astrophysics, Flatiron Institute, 162 5th Ave, New York, NY 10010 }

\date{Last updated 2015 May 22; in original form 2013 September 5}

\pubyear{2020}

\begin{document}
\label{firstpage}
\pagerange{\pageref{firstpage}--\pageref{lastpage}}
\maketitle

\begin{abstract}
The rate of magnetic field diffusion plays an essential role in 
several astrophysical plasma processes. 
It has been demonstrated that
the omnipresent turbulence in astrophysical 
media
induces fast magnetic reconnection, which consequently leads to large-scale magnetic flux diffusion at a 
rate independent of the 
plasma microphysics. 
This process is called ``reconnection diffusion'' (RD)
and allows for the diffusion of fields which are dynamically important. 
The current
theory describing RD 
is
based on incompressible magnetohydrodynamic (MHD) turbulence.
In this  work, we have tested quantitatively 
the predictions of the RD theory 
when magnetic forces are dominant in the turbulence dynamics
(Alfv\'{e}nic Mach number $M_A < 1$).
We employed the \textsc{Pencil Code}
to perform numerical simulations of forced 
MHD 
turbulence, extracting the values of the diffusion coefficient 
$\eta_{RD}$ 
using the Test-Field method. 
Our results are consistent with the RD theory 
($\eta_{RD} \sim M_A^{3}$ for $M_A < 1$) 
when turbulence approaches 
the incompressible limit (sonic Mach number $M_S \lesssim 0.02$), 
while for larger 
$M_S$ 
the diffusion is faster 
($\eta_{RD} \sim M_A^{2}$).
This work shows for the first time simulations of compressible MHD turbulence with 
the suppression of the 
cascade in the direction parallel to the mean magnetic field, 
which is consistent with  incompressible weak turbulence theory.
We also verified that in our simulations the 
energy 
cascading time 
does not follow the scaling with $M_A$ predicted for the weak 
regime, in contradiction 
with the RD theory assumption. 
Our results generally support and expand the RD theory predictions. 
\end{abstract}




\begingroup
\let\clearpage\relax
\endgroup
\newpage

\graphicspath{{./figs/}}

\section{Introduction}

One of the most employed and well known  concepts in magnetohydrodynamic (MHD) theory is the 
magnetic ``frozen-in'' condition introduced by Alfv\'en. When the time scales for  Ohmic 
dissipation of the magnetic fields are much larger than the typical dynamical time scales of 
the flow (the dimensionless parameter characterizing the ratio between these two time scales 
being given by the magnetic Reynolds number $R_M = U L/\eta$,
with $U$ and $L$ the characteristic velocity and scale of the flow, and $\eta$ the 
magnetic diffusivity provided by the Ohmic dissipation), 
one can adopt the ideal MHD approximation. 
It consists in neglecting the resistive term in the magnetic induction equation.
In this limit, it can be demonstrated that the magnetic flux  across a Lagrangian fluid
element is conserved, that is, the magnetic field is perfectly advected by the 
fluid motions in the 
direction 
normal
to the field lines.

The ideal MHD description (and consequently, the frozen-in condition) is usually thought to be 
a good approximation for most astrophysical plasmas, which have in general huge values of $R_M$.
Nevertheless, the frozen in condition when applied, for instance,  to star formation regions 
gives rise to several 
problems  due to observational  and theoretical requirements   
for diffusive magnetic flux transport through the plasma 
(e.g. \citealt{santos-lima_etal_2010, santos-lima_etal_2012, santos-lima_etal_2013, leao_etal_2013, gonzalez-casanova_etal_2016}).
Ambipolar diffusion is usually invoked for breaking the frozen-in condition 
and solving the problem of the magnetic flux transport 
during  star formation 
(see for example \citealt{shu_1983, nishi_etal_1991, ciolek_mouschovias_1993, shu_etal_1994, tassis_mouschovias_2005}).
However, several studies 
revealed weaknesses in this solution 
(\citealt{shu_etal_2006, crutcher_etal_2009, krasnopolsky_etal_2010, krasnopolsky_etal_2011, li_etal_2011}).
A potential solution for this problem in the framework of protoplanetary disk formation has been proposed by 
\cite{machida_etal_2007, machida_etal_2009} based on laminar MHD simulations combined with local Ohmic resistivity 
(see also recent studies on the role of the ambipolar diffusion 
during protostellar disk formation \citealt{wurster_li_2018, guillet_etal_2020, marchand_etal_2020, zhao_etal_2020}). 

We believe that the limitations of the approach that is described above is that the effects of ubiquitous astrophysical turbulence
are disregarded in the aforementioned studies. 
The diffuse interstellar medium 
and molecular clouds are turbulent. 
There are overwhelming observational evidences that support this claim through the measurements of power spectrum of densities in diffuse ISM (see \citealt{armstrong_etal_1995, chepurnov_lazarian_2010}), broadening of the molecular lines (see \citealt{larson_1981}), statistics of velocities (see 
\citealt{lazarian_pogosyan_2000,lazarian_2009} and references therein, 
\citealt{padoan_etal_2009, chepurnov_etal_2010, chepurnov_etal_2015, kandel_etal_2017b, utomo_etal_2019, wolleben_etal_2019, xu_2019, yuen_etal_2019}), 
variations of the Faraday rotation (\citealt{haverkorn_etal_2008, xu_zhang_2016}), 
and
power spectrum of synchrotron fluctuations (\citealt{chepurnov_1998, cho_lazarian_2002b, cho_lazarian_2010}).
The reviews describing the molecular cloud turbulence are presented in 
\cite{mckee_ostriker_2007};
\citet[see also \citealt{elmegreen_scalo_2004}]{maclow_klessen_2004}.

The field of MHD turbulence (see 
\citealt{montgomery_turner_1981, matthaeus_etal_1983, shebalin_etal_1983, higdon_1984})
has seen a rapid progress due 
to both, 
to Solar wind measurements 
(see \citealt{tu_marsch_1995, goldstein_etal_1995, bruno_carbone_2013} for a review), 
theoretical (see \citealt{goldreich_sridhar_1995, lazarian_vishniac_1999, lithwick_goldreich_2001, cho_lazarian_2002, eyink_etal_2011}) 
and numerical progress 
(\citealt{cho_vishniac_2000, maron_goldreich_2001, cho_etal_2002, cho_lazarian_2003, kowal_lazarian_2010, 
federrath_etal_2010, beresnyak_2014};
see also a recent book by \citealt{beresnyak_lazarian_2019}).

As we discuss later, the subject of MHD turbulence is closely related to the processes of magnetic reconnection in turbulent fluid. The model of turbulent reconnection in \cite{lazarian_vishniac_1999} predicts the failure of the traditional flux freezing in highly conducting turbulent fluids. This process does not depend on the rate of ambipolar diffusion and 
motivated a number of our earlier studies (\citealt{lazarian2005, lazarian2011, santos-lima_etal_2010, santos-lima_etal_2012, santos-lima_etal_2013, gonzalez-casanova_etal_2016}) that consider the processes of magnetic flux transport in turbulent fluids. 

The process by which the topology of the magnetic field changes 
depends on whether the fluid is in a laminar or turbulent state. 
 In the presence of turbulence, the motions of the ionized gas produce 
tangling and wandering of the 
magnetic 
field lines which give 
origin to several micro-sites of magnetic reconnection.
This process is 
independent on how small is
the 
Ohmic resistivity which is always present in any real plasma. These reconnection 
micro-sites are continuously formed 
and spread all over the turbulent plasma volume. 
As a consequence, the field lines topology can be modified, and
large-scale
magnetic flux can be transported 
through the gas, implying 
that 
the flux freezing concept is seriously altered (\citealt{lazarian2005, eyink_etal_2011}). 
The speed at which the magnetic flux is transported in such conditions is 
independent of the electric resistivity of the plasma, or the degree of its ionization
but is regulated by the turbulence parameters, as predicted in the theory of fast 
magnetic reconnection introduced  by \cite{lazarian_vishniac_1999}. This theory was
tested numerically in \cite{kowal_etal_2009, kowal_etal_2017, kowal_etal_2020}.
A convincing quantitative numerical study
proving that turbulent reconnection violates flux freezing in MHD turbulence is presented in 
\cite{eyink_etal_2013}.
An extensive body of evidence in favor of turbulence reconnection has been 
collected by now and we refer 
the
reader to 
\cite{lazarian_etal_2020} 
where the modern state of the turbulent reconnection theory and a description of
its tests with 
solar
wind as well as numerical testing with different codes is reviewed. 

The concept of magnetic diffusion via turbulent reconnection 
 ---{\it Reconnection Diffusion} (henceforth RD)--- 
is distinct from the concept of standard turbulent mixing. 
The latter is based on the idea that the field lines are mixed passively by the 
turbulent eddies, without taking into account the effects of the magnetic field on 
the turbulent cascade. 
RD on the other hand,  covers the interesting situation  in which the magnetic 
forces are dynamically important 
(e.g. in the late stages of star formation), and relies on the 
fact that the fast reconnection induced by the MHD turbulence is independent 
of the value of the electric resistivity of the plasma. For many of 
the
astrophysical
applications it is important that the RD is not altered by the effects of ambipolar
drift on the scales where turbulence exists. 

This RD theory predicts that the diffusion coefficient 
$\eta_{RD}$
for large-scale magnetic 
fluxes (i.e., scales larger than the injection or forcing scale of the turbulence) 
depends on the turbulence parameters as follows. In the case of 
super-Alfv\'enic turbulence (that is, when the Alfv\'enic Mach number, 
$M_A = U_{\rm turb} / v_A$ 
[$U_{\rm turb}$ is the turbulent velocity and $v_A$ is the local Alfv\'en velocity] is 
larger than one), it coincides with  the standard turbulent mixing  coefficient,
\begin{equation}
	\eta_{\rm RD} \sim L_{\rm turb} U_{\rm turb}, 
\end{equation}
where $L_{\rm turb}$ and $U_{\rm turb}$ 
are the length and  the  velocity of the turbulence at the injection scale, respectively.
On the other hand, in the regime of sub-Alfv\'enic turbulence ($M_A < 1$), this value is reduced 
by a factor proportional to the third power of $M_A$ 
(\citealt{lazarian_2006, lazarian2011}):
\begin{equation}\label{eq:rd_eta_subalfvenic}
	\eta_{\rm RD} \sim L_{\rm turb} U_{\rm turb} M_A^{3}.
\end{equation}

Therefore, according to the RD theory, the efficiency of the magnetic flux transport strongly depends on the 
local turbulence regime. 
In the context of star formation, the turbulence parameters in scales below sub-parsec (down to dozens of A.U.) can be difficult to infer with precision.
Nonetheless, we reinforce that RD process is always present at some level during all the star formation process.
It is a natural consequence of the ubiquitous presence of turbulence in astrophysical environments
(see for example \citealt{krumholz_mckee_2005, ballesteros-paredes_etal_2007, hennebelle_chabrier_2011, padoan_nordlund_2011, federrath_klessen_2012, federrath_klessen_2013, hull_etal_2017}).
In previous work (\citealt{santos-lima_etal_2010, santos-lima_etal_2012, 
santos-lima_etal_2013, leao_etal_2013, gonzalez-casanova_etal_2016}), we investigated 
numerically the removal of magnetic flux from collapsing turbulent molecular clouds and 
protostellar disks, considering an ``ideal'' MHD approach (i.e., the microscopic magnetic 
dissipation term was not considered explicitly in the induction equation, although an 
effective value is always present due to 
the
numerical discretization of the equations).
We found that the magnetic flux removal by RD  is 
efficient in these systems, 
and helps
the gravitational collapse of the structures (see also \citealt{myers_etal_2013}). 
However, these works focused mostly on the super and  trans-Alfv\'enic regimes of 
the turbulence where the reconnection diffusion coefficient is controlled by Eq.~(1).
\textit{The aim of this work is to test 
quantitatively
the prediction 
of Eq.~(\ref{eq:rd_eta_subalfvenic})
by using 
three-dimensional (3D)  MHD simulations.
It is also the first attempt to generate simulations of 
stationary weak MHD turbulence (the cenario invoked by the RD theory) 
in the presence of finite compressibility, which is more realistic for astrophysical environments. }

This work in organized as follows. In \S2 we present the main predictions of the RD theory. 
The numerical methods and setup for the numerical simulations of this study are described in \S3. 
The results are presented and discussed in \S4 and \S5, respectively. Our major findings are 
finally summarized in \S6.

\section{Basics of reconnection diffusion theory}
\label{sec:rd_theory}

To understand the process of reconnection diffusion we 
present here 
some basic facts of MHD turbulence theory. For simplicity, we consider only the case of incompressible MHD turbulence. 

Traditionally, Alfv\'{e}nic turbulence is described in terms of non-linearly interacting wave packets in Fourier space 
(\citealt{iroshnikov_1963, kraichnan_1965}).
The non-linear cascading rate is given in this case by
\begin{equation}
\tau_{\rm casc}^{-1} \approx \frac{ (\delta u_{\ell} / \ell_{\perp})^2 }{\omega_A}, 
\label{Krai}
\end{equation}
where the angular frequency,
$\omega = V_A / \ell_{\parallel}$, 
and $\ell_{\parallel}$ is the parallel wavelength. 
If the injection velocity $U_L$ is less than the Alfv\'en velocity, the cascade is evolving 
decreasing
only perpendicular 
wavelengths $\ell_{\perp}$.
This is the regime of weak turbulence with the parallel scale being all the time equal to the injection scale and 
$\delta u_{\ell} \sim \ell_{\perp}^{1/2}$ 
(\citealt{lazarian_vishniac_1999} [\citetalias{lazarian_vishniac_1999} hereafter]; 
\citealt{galtier_etal_2000}). However the weak cascade changes its nature at a small scale. Below we explain the nature of this change. 

The theory of strong MHD turbulence was formulated by \citet[henceforth \citetalias{goldreich_sridhar_1995}]{goldreich_sridhar_1995} for transAlfv\'{e}nic turbulence, i.e. $U_L=V_A$. For this turbulence \citetalias{goldreich_sridhar_1995} formulated the condition of the {\it critical balance} that relates the scale of parallel and perpendicular motions, namely, 
$\delta u_{\ell} / \ell_{\perp} \approx V_A / \ell_{\parallel}$. 
This condition means that the parallel scale should change with the decrease of the perpendicular scale. 
 
The transfer to the strong MHD turbulence regime can happen if the turbulence is injected at 
$U_L<V_A$ 
and is weak at its origin. The key to that is the increase of the strength of non-linear interactions with the decrease of perpendicular scale. The transfer to the strong turbulence regime changes the nature of the turbulent motions. In fact, due to fast turbulent reconnection predicted in \citetalias{lazarian_vishniac_1999} the turbulent motions become similar to the hydrodynamic eddies that mix magnetic field perpendicular to the direction of magnetic field. For this eddies it is natural to assume that $\delta u_{\ell}/\ell \sim V_A/\ell_{\|}$, i.e., 
that the rate of turbulent mixing motions is equal to the rate of perturbation propagating along the magnetic field. The difference between the original critical balance in \citetalias{goldreich_sridhar_1995} and the above relation between the perpendicular scale $\ell$ and the parallel scale $\ell_\|$ is that both scales are measured in terms of the magnetic field of the eddies, i.e. {\it local} magnetic field, compared to the wavenumbers $k_\|$ and $k_\bot$ that are given in the global mean 
magnetic field reference system. The concept of local eddies mixing up magnetic field is an essential component 
for the 
understanding 
of 
the RD process. Note
that numerical research does indicate that the critical balance relations are valid only in the local system of reference 
(\citealt{cho_vishniac_2000, maron_goldreich_2001, cho_etal_2002}).
 
The generalization of \citetalias{goldreich_sridhar_1995} theory for subAlfv\'{e}nic turbulence provided in \citetalias{lazarian_vishniac_1999} as well as the analogy between the eddy turbulence in MHD and ordinary
hydrodynamic turbulence helps to understand the nature of the RD. 
The predictions for RD when turbulence 
is in the sub-Alfv\'enic regime (\citealt{lazarian2005, lazarian2011}) can be recovered 
from the statistical calculations 
presented in \cite{eyink_etal_2011}, and are summarized bellow.

The magnetic field is assumed to be diffused by turbulence at a rate similar to the diffusion
of trace particles in the direction perpendicular to the (locally) uniform magnetic field.
A simple statistical analysis can show that the perpendicular diffusivity of trace particles 
due to the turbulent velocity field $\delta \mathbf{u}$ is given by: 
\begin{equation}
	D_{\perp} = \int_{-\infty}^{\infty} dt \langle \delta \mathbf{u}'_{\perp}(0) \cdot \delta \mathbf{u}'_{\perp}(t) \rangle,
\end{equation}
where $\delta \mathbf{u}'(t)$ is the turbulent velocity at the position of the particle at 
time $t$, and the angle brackets, 
$\langle \cdot \rangle$, 
denote an ensemble average
over all the particles, which are assumed to be distributed randomly in space.
Considering turbulence composed by Alfv\'en waves with random phases and turbulence 
with a single scale, 
$\ell$, 
perpendicular to the field lines (and corresponding parallel scale, 
$\ell_{\parallel}$). The expression inside the last integral is 
\begin{equation}
	\langle \delta \mathbf{u}'_{\ell,\perp}(0) \cdot \delta \mathbf{u}'_{\ell,\perp}(t) \rangle \sim \delta u_{\ell}^2 \Re \left\{ \exp \left( i \omega_{A,\ell_{\parallel}}t - |t|/\tau_{\rm dec} \right) \right\},
 \end{equation}
where $\delta u_{\ell}$ is the perpendicular turbulent velocity at scale $\ell$,  $\omega_{A,\ell_{\parallel}} \sim V_A/\ell_{\parallel}$ is the local Alfv\'en frequency, 
and $\tau_{\rm dec}$ is the time-scale 
for decorrelation of the velocity eddies (the correlation is simply assumed to decay exponentially in time). This leads to
\begin{equation} \label{eq:diffusivity_general}
	D_{\perp}(\ell) \sim \delta u_{\ell}^2 \frac{\tau_{\rm dec}}{(\omega_{A,\ell_{\parallel}} \tau_{\rm dec})^2 + 1}.
\end{equation}

We expect  that 
motions at the largest scales
produce the dominant diffusivity. We will assume isotropy at the injection scale, that is, $\ell_{\parallel} \sim \ell$.

When the time-scale for the microscopic diffusivity is  larger than the dynamical time-scales 
of the system (which is the case for most astrophysical environments), the decorrelation time for the velocity eddies 
at the injection scale should be of the order of the energy cascade time for these eddies.
In the regime of weak turbulence (i.e., when the energy cascading time is much longer than the linear wave time
$\tau_{\rm w} \sim \omega_{A, \ell_{\parallel}}^{-1}$), 
we have
$\tau_{\rm casc} \sim (\ell /\delta u_{\ell}) M_A^{-1}$ (\citealt{lazarian_vishniac_1999, galtier_etal_2000}).
Substituting this value in Eq. (\ref{eq:diffusivity_general}), we obtain the RD prediction for the magnetic diffusivity $\eta_{\rm RD}$ (\citealt{lazarian2005}):
\begin{equation} \label{eq:eta_rd_prediction}
	\eta_{\rm RD} \sim D_{\perp} \sim \ell \delta u_{\ell} \min \left( 1, M_A^{3} \right).
\end{equation}

In situations where
the energy cascading time at the injection scale is larger 
than the molecular or numerical viscous time (implying a low effective Reynold's number), 
the decorrelation time of the velocity, 
$\tau_{\rm dec}$,
can be more closely related to the dissipation time, $\tau_{\rm diss} \sim \ell^2 / \nu$,
where $\nu$ is the molecular viscosity.
The magnetic field diffusion driven  by turbulence in this case
will depend on the microscopic diffusion and, if $\tau_{\rm dec} \sim \tau_{\rm diss}$, 
then the diffusivity becomes dependent on the molecular viscosity, 
$\nu$. 

In our discussion we considered only one component of MHD turbulence, namely, Alfv\'en modes and disregarded the slow and fast modes 
(see \citealt{cho_lazarian_2003}). 
This is due to the fact that Alfv\'en modes are the most important in mixing the medium.

\section{Numerical Models}
\label{sec:numerical_models}

In order to test the dependency of the magnetic diffusion coefficient, 
$\eta_{\rm RD}$, 
with the Alfv\'enic Mach number, 
$M_A$,
of the turbulence in the sub-Alfv\'enic regime, we employed three-dimensional MHD numerical 
simulations of forced turbulence in a Cartesian domain, in the presence 
of an external uniform magnetic field of intensity $B_0$ in the $x$-direction. We used the   
\textsc{Pencil Code}~\footnote{http://pencil-code.googlecode.com/} for 
numerically solving the set of compressible, isothermal, 
MHD equations:
\begin{equation}
	\frac{D \ln \rho}{D t} = - \nabla \cdot \mathbf{u},
\end{equation}
\begin{equation} \label{eq:velocity}
	\frac{D {\mathbf u}}{D t} = - c_{s}^{2} \nabla \ln \rho + \frac{1}{\rho} \mathbf{J \times \left( B_0 + B \right)} 
	+ \nu_3 \nabla^6 \mathbf{u} + \mathbf{f},
\end{equation}
\begin{equation} \label{eq:induction}
	\frac{\partial {\mathbf A}}{\partial t} = \mathbf{u \times \left( B_0 + B \right)} + \eta_3 \nabla^6 \mathbf{A}, 
\end{equation}
where $D/Dt = \partial/\partial t + \mathbf{u} \cdot \nabla$ is the lagrangian derivative, 
$\mathbf{A}$ is the magnetic potential vector, $\mathbf{B = \nabla \times A}$ is the magnetic field 
generated by the internal currents, $\mathbf{J = \nabla \times B}/\mu_0$ is the current density, 
$\mu_0$ is the magnetic permeability, 
$\nu_3$ and $\eta_3$ 
are, respectively, the coefficients of hyper-viscosity and magnetic hyper-diffusivity,
$c_s$ is the isothermal sound speed, $\mathbf{u}$ is the velocity,
$\rho$ is the density, and $\mathbf{f}$ 
represents the force responsible for the turbulence injection.
We use the hyper-viscosity and magnetic hyper-diffusivity
schemes with the aim of obtaining a turbulent 
spectra with an extension as large as possible 
for each considered resolution \citep{BO95,HB04}. 
The value of the coefficients were minimized
such that numerical stability is guaranteed.

We have also performed simulations without the hyper-viscosity 
and the magnetic hyper-diffusivity. In these cases, we employed the 
usual viscosity and Ohmic resistivity, namely,  
the terms $\nu_3 \nabla^6 \mathbf{u}$ in eq.~\ref{eq:velocity} 
and $\eta_3 \nabla^6 \mathbf{A}$ in eq.~\ref{eq:induction} 
are replaced by 
$\eta_1 \left\{ \nabla^2 \mathbf{u} + (1/3) \nabla \nabla \cdot \mathbf{u} + 2 \boldsymbol{\mathsf{S}} \cdot \nabla \ln \rho \right\}$ 
and $\eta_1 \nabla^2 \mathbf{A}$, respectively, 
where $\nu_1$ is the constant viscosity, 
$\eta_1$ is the magnetic diffusivity, 
and $\boldsymbol{\mathsf{S}}$ is the rate of strain tensor given by
$S_{ij} = \left\{ (1/2) ( \partial u_i / \partial x_j + \partial u_j / \partial x_i ) - (1/3) \delta_{ij} \nabla \cdot \mathbf{u} \right\}$.
These models are explicitly mentioned in the text whenever they appear.

The turbulence is constantly forced by the increment  of the velocity field with a spectrum of Fourier modes. 
These modes are purely solenoidal, and the phases are randomly changed at every iteration during 
the numerical integration. The turbulence is, 
therefore, nearly statistically homogeneous, 
non-helical, and delta correlated in time. In  \S~\ref{sec:setup} we provide more details about the spectrum 
of the excited velocity modes.

The value of the diffusion coefficient, 
$\eta_{\rm RD}$, 
is extracted from the simulations through the Test-Field method.
It employs a set of passive test magnetic 
fields in order to calculate unambiguously the coefficients of the turbulent mean-fields 
(including 
the diffusion tensor; see the Appendix 
\ref{ap:test-field}
and more details in 
\citealt{brandenburg_subramanian_2005, schrinner_etal_2007, brandenburg_etal_2010}).

\subsection{Setup and parameters}
\label{sec:setup}

\begin{table*}
\centering
\caption{Runs parameters}
\begin{adjustbox}{max width=\textwidth}
\begin{threeparttable}
\begin{tabular}{l r r c l l c c c l c}
\hline\noalign{\smallskip}
runs set 
& $L_{\parallel}$x$L_{\perp}$
& res.
& $M_S$\tnote{a} 
& $M_A$\tnote{b} 
& $v_{rms} / v_0$ 
& forcing
& $k_{\parallel}L / 2\pi$ 
& $k_{\perp}L / 2\pi$\tnote{c} 
& $[\tilde{t}_0, \tilde{t}_1]$\tnote{d} 
& $\tilde{\nu}_3$, $\tilde{\eta}_3$\tnote{e} \\ 
\hline\noalign{\smallskip}
	16L-Ms0.32-A  & 16Lx1L  & $2048,128^2$  & $0.32$  & $0.8$, $0.57$,         & $1.10$, $1.18$,          & A   & $[0,4]$  & $[3,4]$  & $[3,8]$              & $1.8 \times 10^{-9}$ \\
		      &         &               &         & $0.4$, $0.2$           & $1.18$, $1.22$           &     &          &          &                      &                      \\
		      &         &               &         & $0.1$                  & $1.33$                   &     &          &          &                      &                      \\
\hline
	16L-Ms0.08    & 16Lx1L  & $2048,128^2$  & $0.08$  & $0.8$, $0.4$,          & $1.09$, $1.17$,          & A   & $[0,4]$  & $[3,4]$  & $[3,8]$, $[4,9]$     & $1.8 \times 10^{-9}$ \\
		      &         &               &         & $0.28$, $0.2$          & $1.11$, $1.10$           &     &          &          & $[7,12]$, $[7,12]$   &                      \\
		      &         &               &         & $0.1$                  & $1.11$                   &     &          &          & $[9,14]$             &                      \\
\hline
	16L-Ms0.02-A  & 16Lx1L  & $2048,128^2$  & $0.02$  & $0.8$, $0.4$,          & $1.08$, $1.17$,          & A   & $[0,4]$  & $[3,4]$  & $[3,8]$, $[4,9]$     & $1.8 \times 10^{-9}$ \\
		      &         &               &         & $0.28$, $0.2$          & $1.27$, $1.04$           &     &          &          & $[4,9]$, $[13,18]$   &                      \\
		      &         &               &         & $0.1$                  & $1.04$                   &     &          &          & $[15,20]$            &                      \\
\hline
	8L-Ms0.02-A   & 8Lx1L   & $1024,128^2$  & $0.02$  & $0.8$, $0.4$,          & $1.09$, $1.17$,          & A   & $[0,4]$  & $[3,4]$  & $[3,8]$, $[4,9]$     & $1.8 \times 10^{-9}$ \\
		      &         &               &         & $0.2$, $0.1$           & $0.99$, $1.01$           &     &          &          & $[13,18]$, $[15,20]$ &                      \\
\hline
	4L-Ms0.02-A   & 4Lx1L   & $512,128^2$   & $0.02$  & $0.8$, $0.4$,          & $1.10$, $1.19$,          & A   & $[0,4]$  & $[3,4]$  & $[3,8]$, $[4,9]$     & $1.8 \times 10^{-9}$ \\
		      &         &               &         & $0.2$, $0.1$           & $0.94$, $1.01$           &     &          &          & $[13,18]$, $[15,20]$ &                      \\
\hline
	1L-Ms0.02-A   & 1Lx1L   & $128,128^2$   & $0.02$  & $0.8$, $0.4$,          & $1.22$, $1.19$,          & A   & $[0,4]$  & $[3,4]$  & $[3,8]$, $[4,9]$     & $1.8 \times 10^{-9}$ \\
		      &         &               &         & $0.2$, $0.1$           & $1.04$, $1.11$           &     &          &          & $[7,12]$, $[7,12]$   &                      \\
\hline
	8L-Ms0.02-Ab  & 8Lx1L   & $1024,128^2$  & $0.02$  & $0.8$, $0.4$,          & $1.07$, $1.10$,          & Ab  & $-$      & $[4,5]$  & $[3,8]$, $[4,9]$     & $1.8 \times 10^{-9}$ \\
		      &         &               &         & $0.2$, $0.1$           & $1.09$, $0.93$           &     &          &          & $[10,15]$, $[15,20]$ &                      \\
\hline
	8L-Ms0.02-I   & 8Lx1L   & $1024,128^2$  & $0.02$  & $0.8$, $0.4$,          & $1.07$, $1.11$,          & I   & $-$      & $[3,4]$  & $[3,8]$, $[4,9]$     & $1.8 \times 10^{-9}$ \\
		      &         &               &         & $0.2$, $0.1$           & $1.19$, $1.19$           &     &          &          & $[4,9]$, $[4,9]$     &                      \\
\hline
	16L-Ms0.02-low-A & 16Lx1L  & $1024,64^2$   & $0.02$  & $0.8$, $0.4$,          & $1.02$, $1.12$,          & A   & $[0,4]$  & $[3,4]$  & $[11,15]$            & $6.2 \times 10^{-8}$ \\
		         &         &               &         & $0.2$, $0.14$,         & $1.31$, $1.43$           &     &          &          &                      &                      \\
		         &         &               &         & $0.1$, $0.05$          & $1.44$, $1.40$           &     &          &          &                      &                      \\
		         &         &               &         & $0.025$                & $1.32$                   &     &          &          &                      &                      \\
\hline
16L-Ms0.02-low-A-diff2 & 
16Lx1L  & 
$1024,64^2$   & 
$0.02$  & 
$0.8$, $0.4$,          & 
$1.18$, $1.28$,          & 
A   & 
$[0,4]$  & 
$[3,4]$   & 
$[4,8]$            & 
$(1.2 \times 10^{-2})$\tnote{f} \\
		         &         &               &         & 
$0.2$, $0.1$,         & 
$1.37$, $1.43$           &     
&          &          &                      &                      \\
\hline
16Lx2L-Ms0.02-low-A & 
16Lx2L  & 
$1024,128^2$  & 
$0.02$  & 
$0.2$          & 
$1.30$          & 
A   &
$[0,4]$   & 
$[3,4]$            & 
$[11,15]$            & 
$6.2 \times 10^{-8}$ \\
\hline
16L-Ms0.02-hi-A & 
16Lx1L  & 
$2048,256^2$  & 
$0.02$  & 
$0.2$          & 
$0.90$          & 
A   &
$[0,4]$   & 
$[3,4]$            & 
$[3,5]$\tnote{g}            & 
$(5.5 \times 10^{-11})$\tnote{h} \\
\hline
\end{tabular}	
\begin{tablenotes}
\vskip 0.1in
\item[a] $M_S \equiv v_0 / c_s$ is the approximate sonic Mach number of the simulations.
\item[b] $M_A \equiv v_0 / v_{A,0}$ is the approximate Alfvénic Mach number of the simulations.
\item[c] For the models with forcings `Ab' and `I', this column shows the absolute values  of the vector $\mathbf{k}$ of the forced modes.
\item[d] $[\tilde{t}_0, \tilde{t}_1]$ is the time interval used for the averages in time, in units of 
$\ell_{\perp}/v_0$.
\item[e] $\tilde{\nu}_3$, $\tilde{\eta}_3$ are the hyper-viscosity and hyper-resistivity in units of 
$\ell_{\perp}^5 v_0$.
\item[f] No hyper-viscosity or hyper-resistivity 
were
used in these runs. 
This column shows the values of $\tilde{\nu}_1$ and $\tilde{\eta}_1$, the viscosity and magnetic diffusivity, 
respectively, in units of $\ell_{\perp} v_0$. See Section~\ref{sec:numerical_models} for more details. 
\item[g] This simulation 
used as initial 
condition
a previously evolved simulation with resolution $2048,128^2$. 
\item[h] Due to the anisotropic resolution in this simulation,
these values of $\tilde{\nu}_3$ and $\tilde{\eta}_3$ refer to the terms in the 
hyper-viscosity and hyper-resistivity 
containing the derivatives in the directions perpendicular to the 
mean magnetic field. The values of $\tilde{\nu}_3$ and $\tilde{\eta}_3$
used in the terms containing the derivatives in the parallel direction 
are $1.8 \times 10^{-9}$. 
\end{tablenotes}
\label{tab:runs_params_1}
\end{threeparttable}
\end{adjustbox}
\end{table*}

The Reconnection Diffusion theory (\S~\ref{sec:rd_theory}) is formulated in the 
incompressible limit, assuming that when sub-Alfv\'enic turbulence is forced isotropically in the 
presence of an uniform magnetic field (at least locally), it will develop a cascade in the regime 
predicted by the weak turbulence theory.
Therefore, we restricted our simulations setup and parameters to the subsonic regime
(yet using finite sound speed $c_s$ in the simulations) and favoring conditions under 
which weak turbulence could develop. 

Table~\ref{tab:runs_params_1} 
lists the parameters employed in the simulations 
presented in the next Section. 
For each set of simulations (identified by its name in the first column of Table~\ref{tab:runs_params_1}),
we spanned a range of values of $M_A$ by changing the magnetic field strength $B_0$ and 
keeping the rms value of the turbulence velocity, $v_{\rm rms}$, constant, near to a fixed 
reference value, 
$v_0$. We do not control directly
$v_{\rm rms}$ in our simulations, but instead, the amplitude of the forcing. 

The turbulent diffusivity is controlled by the largest scale motions of 
the system (the motions at the injection scale), where the universal laws 
of the inertial range are not formally 
valid, and could be affected by the details of the forcing mechanism. At the same time, 
the forcing can also determine the MHD turbulence regime. Because this work 
focuses on 
the subsonic 
turbulence, we employed forcing schemes purely solenoidal, and with approximately the same coherence 
length in the directions parallel and perpendicular to the mean magnetic field.
The forcing was chosen delta correlated in time for two reasons: first, if the cascading time 
follows the weak turbulence theory, the correlation of the waves at the injection scale 
should persist 
for a time similar to the cascading time, even if the forcing is generating low energy random 
waves continuously.
Second, forcing with time correlation proportional to $M_A^{-1}$ (e.g., \citealt{alexakis2011}) 
must
induce Alfv\'en waves with approximately the same decorrelation time, which could determine 
trivially the diffusion coefficient.

We employed three forcing schemes, which differ in the discrete spectrum of 
the velocity modes excited. In one of them, all the modes inside a spherical 
shell in the $\mathbf{k}$-space are excited with the same amplitude. 
This $\mathbf{k}$-isotropic distribution of amplitudes is the most usual choice 
in numerical simulations of forced turbulence.
The models using this scheme 
are identified by `I' in the column ``forcing'' of Table~\ref{tab:runs_params_1}.
Alternatively, in order to constrain the parallel and perpendicular injection 
scales to well defined values, 
$\ell_{\parallel}$ and $\ell_{\perp}$,
respectively, 
and at the same time avoid the forcing of 
purely 2D modes
(i.e., with $k_{\parallel} = 0$) 
as well as waves of large wavelength 
in the direction parallel to the 
imposed
uniform magnetic field, we also used an {\bf k}-anisotropic scheme 
(identified by `A' in Table~\ref{tab:runs_params_1}). 
It forces
all the waves with the components $k_{\perp}$ 
and $k_{\parallel}$ 
inside a cylindrical shell in the $\mathbf{k}$-space. The amplitude of the 
spectrum is modulated by a factor $\propto k_{\parallel}^2$.
The third forcing scheme we used 
excites all the modes inside a spherical shell in the $\mathbf{k}$-space, but 
modulates
the amplitudes by 
a 
factor 
$\propto \sin (2 \theta$), 
where $\cos \theta = k_{\parallel} / k$. 
This last scheme is identified by `Ab' 
in Table~\ref{tab:runs_params_1}.
The `A' and `Ab' schemes favor 
distributions with reduced amplitude for the wavevectors corresponding to wavelengths
far from the fixed injection scale.
The forcing scheme `I', on the other side,
generates waves elongated in both parallel and perpendicular directions.
We 
emphasize that, statistically, 
all the three schemes force velocity fluctuations 
in an (nearly) isotropic form in the physical space.

The injection scales are indicated in the columns $k_{\parallel}L / 2\pi$ and 
$k_{\perp}L / 2\pi$, 
where ${\rm L}$ is the shortest side of the domain, 
perpendicular to the mean uniform magnetic field, 
in Table~\ref{tab:runs_params_1}
(except for the runs set 16Lx2L-Ms0.02-low-A, where this side has length 2L). 
These values were chosen in order to maximize the turbulence 
inertial range. 
Note, however, that although the separation of scales between the largest 
turbulence eddies and the mode employed in the test fields 
($k_{\perp, {\rm tf}} {\rm L} / 2 \pi = k_{z, {\rm tf}} {\rm L} / 2 \pi = 1$; see Appendix \ref{ap:test-field}) 
is still limited.

Unlike real extended astrophysical environments, the finiteness of the 
computational box introduces effects on the wave turbulence 
due to insufficient density of large-scale modes represented in the 
discrete Fourier space. A theoretical constraint on 
the validity of the classical weak turbulence regime in a finite 
box domain, which was derived both under the Reduced MHD approximation 
and the assumption of $k_{\perp} \gg k_{\parallel}$ (\citealt{nazarenko2007}), 
is approximately given by
\begin{equation} \label{eq:wwt_validity}
	\sqrt{ \frac{\ell_{\parallel}}{L_{\parallel}} } \ll \frac{\delta u_{\ell}}{B_0} \frac{\ell_{\parallel}}{\ell_{\perp}} \ll 1,
\end{equation}
where $\delta u_\ell$ is the velocity amplitude corresponding to 
the mode with scales $\ell_{\parallel}$ and $\ell_{\perp}$. 
Therefore, to fullfill this condition at the injection scale where 
$\ell_{\parallel} \approx \ell_{\perp}$, it is needed to ensure
$\sqrt{ \ell / L_{\parallel} } \ll M_A \ll 1$.

Another effect due to the finite box size is the 2D ``enslaving'' of the 3D MHD turbulence if
\begin{equation} \label{eq:2d_slaving}
	\frac{\ell_{\perp}}{L_{\perp}}
	\sqrt{ \frac{\ell_{\parallel}}{L_{\parallel}} } 
	\gg \frac{\delta u_{\ell}}{B_0} \frac{\ell_{\parallel}}{\ell_{\perp}},
\end{equation}
(\citealt{nazarenko2007}), which at the injection scales becomes 
$( \ell / L_{\perp} ) \sqrt{ \ell / L_{\parallel} } \ll M_A$.
It should be noted that if the leftmost inequality in Eq. (\ref{eq:wwt_validity}) is satisfied, then the inequality in Eq. (\ref{eq:2d_slaving}) is automatically false ($\ell/L_{\perp} < 1$).

To evaluate the finite size effects of the computational domain on the RD coefficient we ran simulations 
with four different domain sizes in the direction parallel to the mean magnetic field ($x$-direction): $L_{\parallel} =$ 16L, 8L, 4L, and 1L, 
keeping fixed the domain size perpendicular L$_{\perp} =$ 1L.
To verify a possible influence of the perpendicular domain size on the RD coefficients, we ran a comparative model with L$_{\perp} =$ 2L, keeping $L_{\parallel} =$ 16L 
(runs set 16Lx2L-Ms0.02-low-A). 

In the results presented in the next Section 
the analysis of the simulations 
is performed
after 
the turbulence has reached the statistically stationary state. 
The analyses average quantities between times $\tilde{t}_0$ and $\tilde{t}_1$ shown in 
Table~\ref{tab:runs_params_1}. 
Four complete snapshots 
with equal time separation 
are extracted from the simulations during this time interval
(the only exception is the model 16Lx1L-Ms0.02-hi-A, which has only two complete snapshots). 

\begin{figure*}
\begin{tabular}{c}
	\input{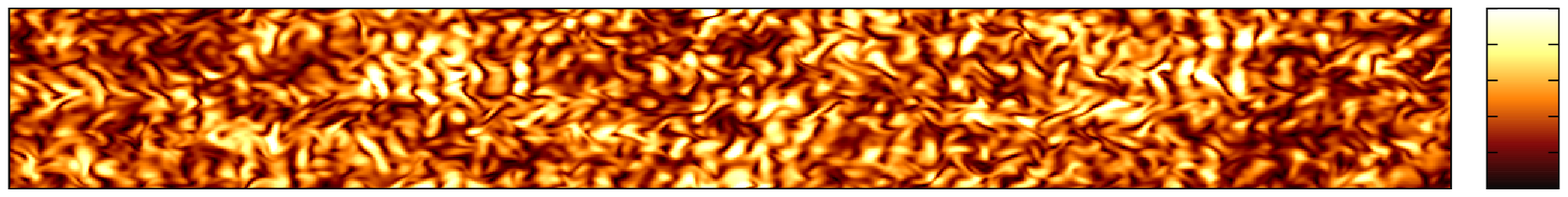} \\
	\input{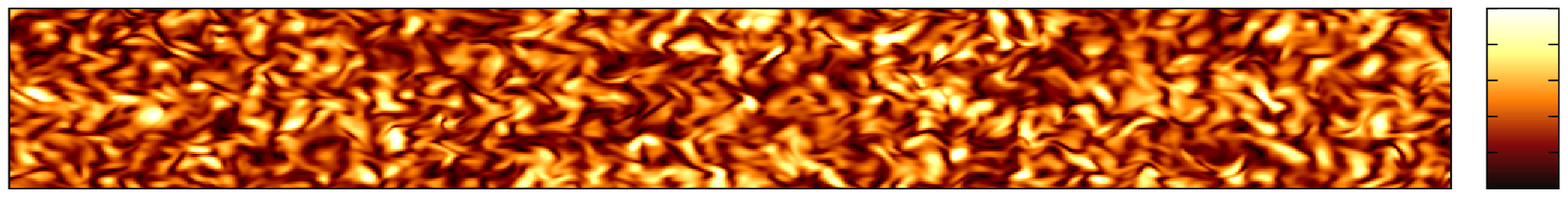} \\
	\input{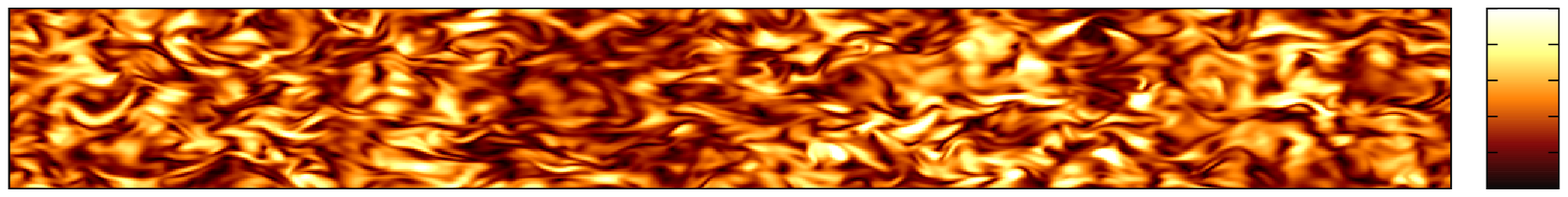} \\
\end{tabular}
	\caption{Central slice (xy-plane) showing the velocity modulus distribution at the final time of the simulations. 
	Models with identical parameters but different forcing distributions in the k-space are compared. From top to bottom: 
	A-forcing, Ab-forcing, and I-forcing. All simulations have the same Alfv\'{e}nic Mach number $M_{A} \equiv v_{0}/v_{A,0} = 0.4$ 
	and sonic Mach number $M_{S} = 0.02$.
        See Table~\ref{tab:runs_params_1} for the complete description of the simulations parameters. }
\label{fig:maps_vel_ma0p4}
\end{figure*}

\begin{figure*}
\begin{tabular}{c}
	\input{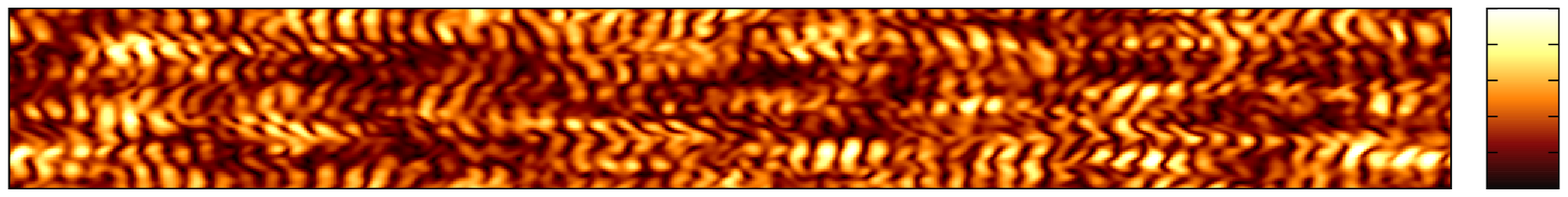} \\
	\input{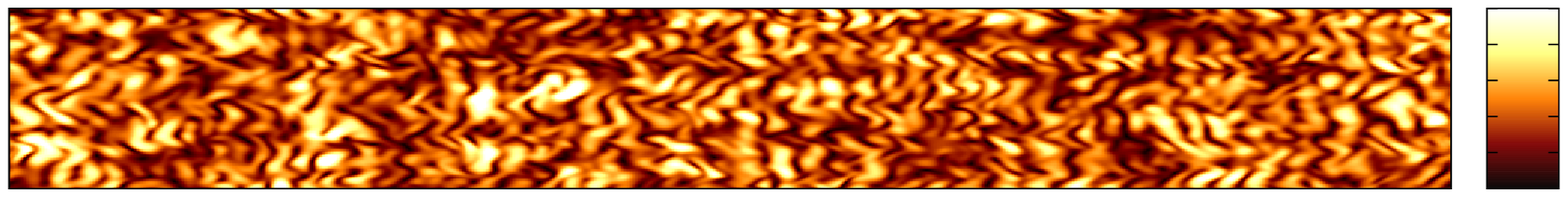} \\
	\input{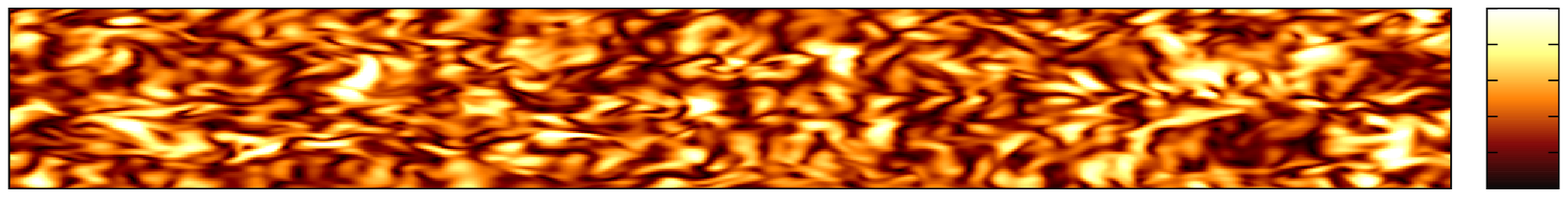} \\
\end{tabular}	
	\caption{Same as Figure~\ref{fig:maps_vel_ma0p4}, but for simulations with the Alfv\'{e}nic Mach number $M_{A} \equiv v_{0}/v_{A,0} = 0.2$. }
\label{fig:maps_vel_ma0p2}
\end{figure*}

Figure~\ref{fig:maps_vel_ma0p4} compares the distribution of the velocity modulus 
on the central xy-plane for three selected runs from Table~\ref{tab:runs_params_1} at time $\tilde{t}_1$ 
(see Table~\ref{tab:runs_params_1}). 
The three models differ only by the forcing scheme: A (top), Ab (middle), and I (bottom). 
All the runs in Figure~\ref{fig:maps_vel_ma0p4} have $M_A = 0.4$. 
Figure~\ref{fig:maps_vel_ma0p2} 
depicts the same quantity for simulations that
have $M_A = 0.2$.
Observe that the differences between the 
turbulent structures generated by the 
different forcings become more pronounced for smaller values of 
$M_A$ (Figure~\ref{fig:maps_vel_ma0p2}). 
For $M_A = 0.8$ 
(not shown here for compactness),
the velocity distribution resulting from the three forcing schemes is indistinguishable. 

Below, we define the 2D power spectrum of the turbulence $E(k_{\parallel},k_{\perp})$ 
in terms of the velocity $\mathbf{u}$ and 
magnetic field $\mathbf{B}$ as
\begin{equation}
	E(k_{\parallel},k_{\perp}) = \sum_{\mathbf{k}'} \left( \frac{1}{2} \rho_0 \mathbf{u_{k'}^*} \cdot \mathbf{u_{k'}} + \frac{1}{8 \pi} \mathbf{B_{k'}^*} \cdot \mathbf{B_{k'}} \right),
\end{equation}
where ${\mathbf u_k} = \mathcal{F}_{\mathbf k} \left\{ \mathbf{u} \right\}$ and 
${\mathbf B_k} = \mathcal{F}_{\mathbf k} \left\{ \mathbf{B} \right\}$ are the 
$\mathbf{k} = (\mathbf{k}_{\parallel} + \mathbf{k_{\perp}})$ components of the 
discrete Fourier transform of $\mathbf{u}$ and $\mathbf{B}$,
respectively, $\mathbf{k}_{\parallel} = (\mathbf{k \cdot B}_0) \mathbf{B}_0 / B_0^2$, 
$\rho_0$ is the mean density,
and $k_{\parallel,\perp} = |\mathbf{k}_{\parallel,\perp}|$.
The superscript $^*$ means 
the complex conjugate, and the sum extends for all the 
discrete
modes,
$\mathbf{k'}$,
with components in the interval $k_{\parallel} \le | \mathbf{k}_{\parallel}' | < (k_{\parallel} + 1)$,
and $k_{\perp} \le | \mathbf{k_{\perp}'} | < (k_{\perp} + 1)$. 
The 1D power spectrum $E(k_{\perp})$ is defined by
\begin{equation}
	E(k_{\perp}) = \sum_{k_{\parallel}=0}^{k_{\parallel,\max}} E(k_{\parallel},k_{\perp}),
\end{equation}
such that
\begin{equation}
	E_{\rm turb} = \frac{1}{2} \rho_0 \langle \mathbf{u}^2 \rangle + \frac{1}{8 \pi} \langle \mathbf{B}^2 \rangle = 
 \sum_{k_{\perp}=1}^{k_{\perp,\max}} E(k_{\perp}),
\end{equation}
is the total turbulent energy in the system, and the brackets,
$\langle \cdot \rangle$,
represent average in space.

The transfer spectrum is obtained from the following procedure: 
we multiply $\mathbf{u^*_k}$ to the Fourier transform of the momentum equation 
(assuming the incompressible limit by ignoring the density variations), and we add the  
Fourier transform of the induction equation multiplied by $\mathbf{B^*_k/(4 \pi)}$.
We denote by $T_{\mathbf k}$ 
the time derivative of $( \rho_0 \mathbf{u_{k}^*} \cdot \mathbf{u_{k}}/2 + \mathbf{B_{k}^*} \cdot \mathbf{B_{k}}/8 \pi)$, when we neglect the forcing and dissipation terms:
\begin{eqnarray}
	T_{\mathbf k} &=& 
\Re \left\{ \rho_0 \mathbf{u^*_k} \cdot \mathcal{F}_{\mathbf k} \left[ \mathbf{ \left( u \cdot \nabla \right) u - \frac{1}{4 \pi} \left( \nabla \times B \right) \times B } \right] \right\} \nonumber \\
	&& - 
\frac{1}{4 \pi} 
\Re \left\{ \mathbf{B^*_k} \cdot \mathcal{F}_{\mathbf k} \left[ \mathbf{ \nabla \times \left( u \times B \right) } \right] \right\}, 
\end{eqnarray}
with $\Re$ denoting the real part of. 

The perpendicular transfer spectrum $T(k_{\perp})$ is then
defined by
\begin{equation}
	T(k_{\perp}) = \sum_{\mathbf{k'}} T_\mathbf{k'},
\end{equation}
with the sum extending 
over all the modes 
$\mathbf{k'}$,
with the perpendicular components, 
$\mathbf{k_{\perp}'}$,
in the interval  
$0 \le | \mathbf{k_{\perp}'} | < (k_{\perp} + 1)$
(see, for example, \citealt{alexakis_etal_2007, alexakis2011})
. The turbulence 
energy transfer is given by the maximum value of $T(k_{\perp})$,
\begin{equation}
	T_{\rm turb} = \max \left\{ T(k_{\perp}) \right\}.
\end{equation}

As we are going to see in Section~\ref{sec:results_forcing}, 
Figure~\ref{fig:maps_ps_F_ma} 
shows the normalized 2D power spectrum distribution
$E(k_{\parallel},k_{\perp})$ for the same models 
presented in Figures~\ref{fig:maps_vel_ma0p4} and~\ref{fig:maps_vel_ma0p2}, in the left and right column, respectively.
By observing the distribution of modes with the highest values of energy (around $0 < k_{\parallel}L / 2\pi, k_{\perp}L / 2\pi < 4$),
we can 
qualitatively assess 
the differences between the three different forcing schemes.

Observe that in the above definitions of energy and transfer spectrum, we are neglecting any density fluctuations. 
We choose to do so, in order to simplify the analysis 
and because we expect the Alfvén modes to dominate the turbulence spectrum
for the subsonic simulations presented in this work.

\section{Results}
\label{sec:results}

\subsection{Compressibility and domain size effects}

\begin{figure*}
\begin{tabular}{c c}
	\input{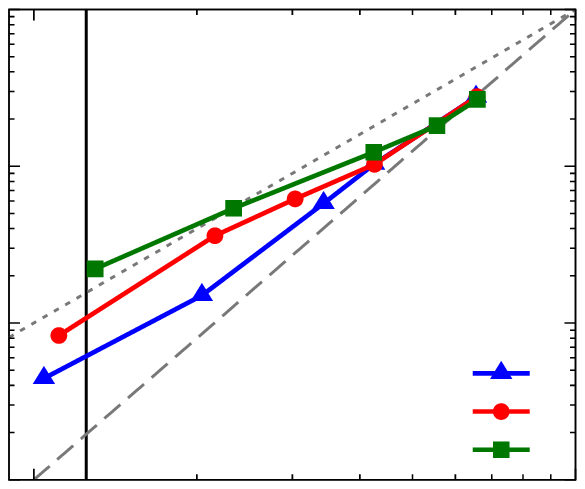} &
	\input{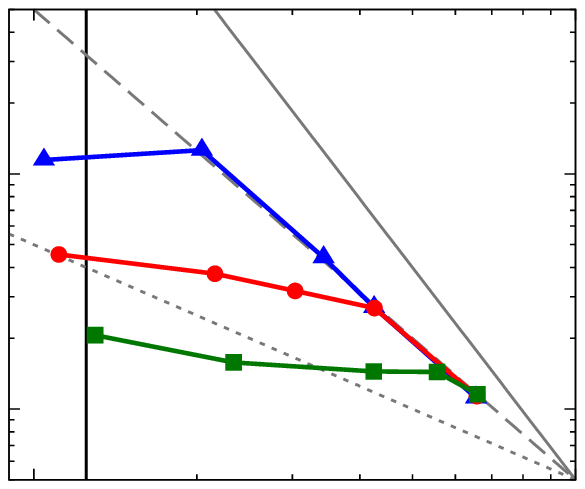} \\
\end{tabular}	
	\caption{Magnetic diffusion coefficients $\eta_{\rm tf}$ measured by the test-field method (left)
	and the energy transfer time $\tau_{\rm ener} \equiv E_{\rm turb} / T_{\rm turb}$ (right),
	as a function of the Alfv\'{e}nic Mach number $M_A = v_{\rm rms}/\langle v_{A} \rangle$. 
	Simulations with the same A-forcing, the domain sizes, but different sonic Mach number $M_S = v_{\rm rms}/c_s$ are compared.
	Each point corresponds to one run in Table~\ref{tab:runs_params_1}. 
	For the parameters used in these simulations the vertical solid line indicates the
	lower limit of $M_A$ given by Eq.~\ref{eq:wwt_validity} at the injection scale $\ell$.}
\label{fig:ma_eta_tener_cs}
\end{figure*}

\begin{figure*}
\begin{tabular}{c c}
	\input{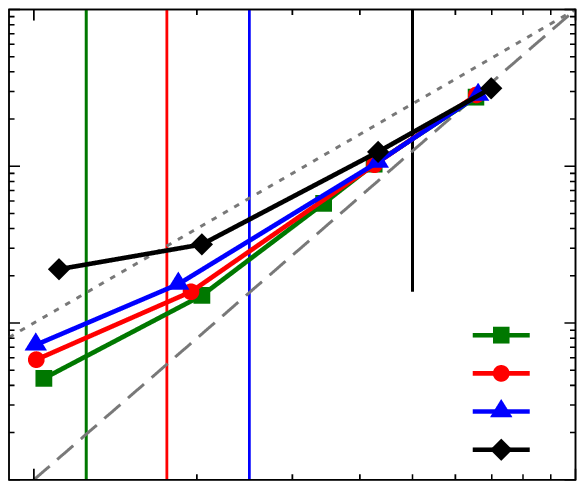} &
	\input{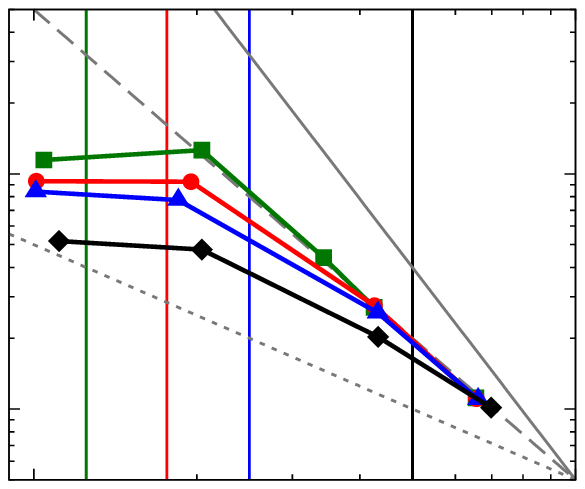} \\
\end{tabular}	
	\caption{Magnetic diffusion coefficients $\eta_{\rm tf}$ measured by the test-field method (left)
	and the energy transfer time $\tau_{\rm ener} \equiv E_{\rm turb} / T_{\rm turb}$ (right),
	as a function of the Alfv\'{e}nic Mach number $M_A = v_{\rm rms}/\langle v_{A} \rangle$. 
	Simulations with the same A-forcing, the same sonic Mach number $M_S = 0.02$, but different domain sizes are compared. 
	Each point corresponds to one run in Table~\ref{tab:runs_params_1}. 
	For each domain size (corresponding to different  color curves), 
	a vertical solid line indicates the
	lower limit of $M_A$ given by Eq.~\ref{eq:wwt_validity} at the injection scale $\ell$.}
\label{fig:ma_eta_tener_lx}
\end{figure*}

Left panels of Figures~\ref{fig:ma_eta_tener_cs} and \ref{fig:ma_eta_tener_lx} show 
the dependence of 
$\eta_{\rm tf}$, computed with the test-field method,  with the $M_A$ resulting in the simulations 
(see for example \citealt{schrinner_etal_2005, brandenburg_etal_2008}).
Each point in the curves  corresponds to one model  in Table~\ref{tab:runs_params_1} with anisotropic  forcing (A), 
and points with the same shape and color correspond to models with the same sonic Mach number in Figure~\ref{fig:ma_eta_tener_cs}, 
and same box size in Figure~\ref{fig:ma_eta_tener_lx}.
The values of the diffusivity are normalized by the estimate of the hydrodynamical turbulent diffusivity 
$\eta_{\rm hyd} = (1/3) \ell v_{\rm rms}$, where $\ell$ is the injection scale of the simulation 
(corresponding to the minimum wavenumber $k_{\perp} L / 2\pi$ indicated in Table~\ref{tab:runs_params_1}). 
The Alfv\'{e}nic Mach number measured from the simulations is defined by,
$M_A = v_{\rm rms} / \langle v_{A} \rangle$, 
where $\langle v_{A} \rangle = \langle B / \left( 4 \pi \rho \right)^{1/2} \rangle$ is the 
Alfv\'{e}n velocity averaged in the domain. The quantities $v_{\rm rms}$, $\langle v_{A} \rangle$, and $\eta_{\rm tf}$ 
represent time averaged values of the respective space averaged quantities. 

The left panel of Figure~\ref{fig:ma_eta_tener_cs} compares models with the same domain size 16Lx1L, 
but three different values of sonic Mach number, $M_S = v_0/c_s$ (where $c_s$ is the isothermal sound speed of the simulation). 
For  $M_A$ larger than a certain value $M_A^*$ ($M_A^* \approx 0.5, 0.4,$ and $0.2$ for 
$M_S = 0.32, 0.08,$ and $0.02$, respectively), we observe the approximate relation $\eta_{\rm tf}/\eta_{\rm hyd} \sim M_A^3$. 
For $M_A < M_A^*$, this power-law dependence with $M_A$ seems to change 
asymptotically
to $\eta_{\rm tf} / \eta_{\rm hyd} \sim M_A^2$, with the constant of proportionality increasing with $M_S$.
For the models with the lowest Mach number ($M_S = 0.02$) the value of $M_A^*$ is closer to the 
theoretical lower limit for validity of the classical weak turbulence (Equation~\ref{eq:wwt_validity}).
This limit 
is indicated by the vertical solid line in Figure~\ref{fig:ma_eta_tener_cs}.

\begin{figure*}
\begin{tabular}{c c}
	\input{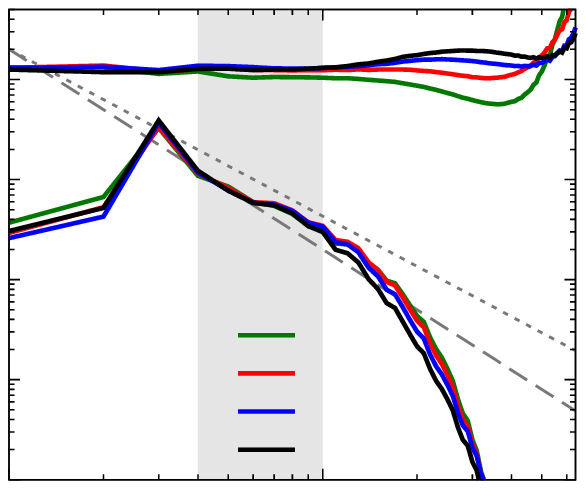} &
	\input{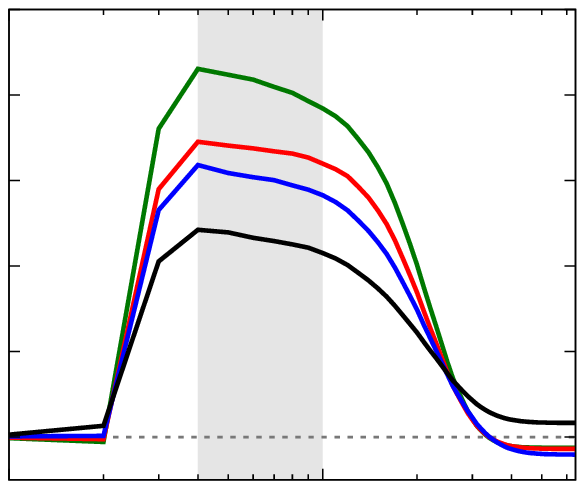} \\
	\input{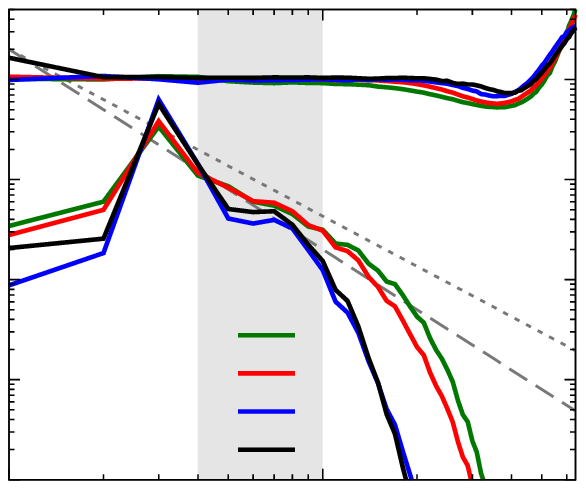} &
	\input{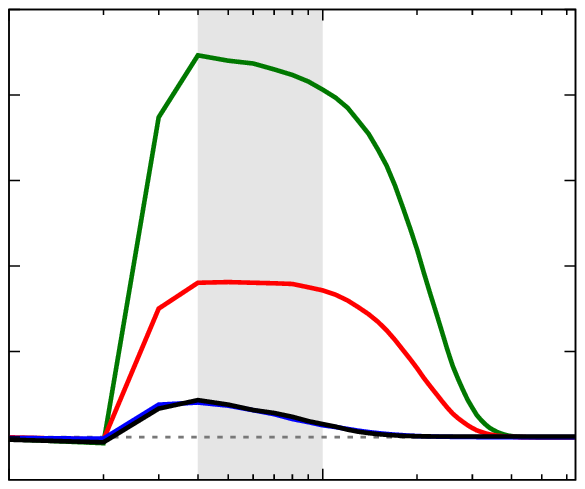} \\
	\input{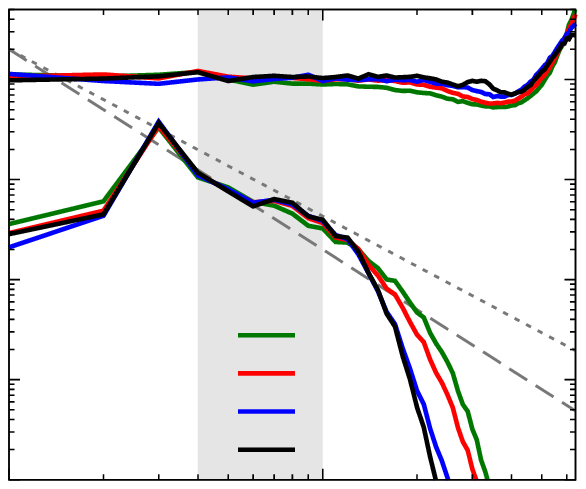} &
	\input{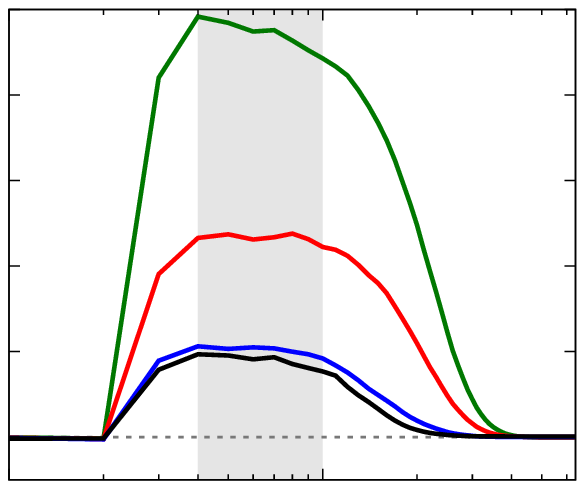} \\
\end{tabular}	
	\caption{The energy spectrum $E(k_{\perp})$ 
	(left column) and the energy transfer spectrum $T (k_{\perp})$ (right panel). 
	The ratio between the 
        velocity 
        and the magnetic 
        power spectra is also shown in the energy spectrum plots.
	In each panel, models from Table~\ref{tab:runs_params_1} with different Alfv\'enic Mach numbers $M_A$ are represented by curves with different colors.
	Top: models  with sonic Mach number $M_S = 0.32$ and domain size 16Lx1L; middle: models with $M_S = 0.02$ and domain size 16Lx1L; bottom: models with $M_S = 0.02$ and domain size 1Lx1L.
	The power laws $\propto k_{\perp}^{-5/3}$ and $\propto k^{-2}$ are also depicted for comparison in the left panels.
	The gray area covers the wavenumbers range $4 < k_{\perp}L / 2 \pi < 10$ for which the transfer spectrum is approximately constant and close to its maximum value.}
\label{fig:ps_ts_1}
\end{figure*}

The turbulent diffusion is naturally dominated by the motions of the injection scales, 
and the decorrelation time of the velocity fields at these scales can be expected to be directly related 
to the energy transfer time shown in the right panels of Figures~\ref{fig:ma_eta_tener_cs} 
and \ref{fig:ma_eta_tener_lx}. 
If the break of the dependence $\eta_{\rm tf}/\eta_{\rm hyd} \sim M_A^3$ around some $M_A^*$ 
is caused by a change in the regime of the turbulence cascade (see Equation~\ref{eq:diffusivity_general}), 
the behavior of the energy transfer time 
$\tau_{\rm ener}$ (i.e., the time-scale for the energy in the injection scale to cascade to smaller scales)
could also occur around the same values of $M_A^*$. 
We estimated $\tau_{\rm ener}$ 
dividing the total turbulence energy $E_{\rm turb} = E_{v} + E_{b}$, where $E_{v}$ and $E_{b}$ are the 
kinetic and magnetic energies (in the incompressible limit), by the maximum value 
of the energy transfer spectrum, $T_{\rm turb}$ 
(which is approximately the energy transfer rate at the 
injection scale or the turbulence injection power; 
see below the description 
of the energy transfer 
spectrum, 
$T (k_{\perp})$,
in the right columns of Figures~\ref{fig:ps_ts_1} 
and~\ref{fig:ps_ts_2}).
Right panel of Figure~\ref{fig:ma_eta_tener_cs} shows the dependence of $\tau_{\rm ener}$ (normalized 
by the estimate of the  non-linear turbulence time $\tau_{\rm nl} = \ell / v_{\rm rms}$ at the 
injection scale), as a function of $M_A$, for the same models 
shown in the left panel. In fact, 
$\tau_{\rm ener} / \tau_{\rm nl}$ follows approximately a well defined 
power law in $M_A$ for $M_A > M_A^*$, and 
becomes flatter for smaller values of $M_A$. 
The 
resulting
power law 
does not coincide with that in the inertial range predicted for the weak 
turbulence regime, 
$\tau_{\rm ener} / \tau_{\rm nl} \sim M_A^{-1}$. Instead, we obtain 
a dependency $\sim M_A^{-2}$. 

Models with the same sonic Mach number,
$M_S = 0.02$, but different domain sizes are compared in the left
panel of Figure~\ref{fig:ma_eta_tener_lx}. 
The shorter the parallel extension $L_{\parallel}$ of the domain, the larger the departure from 
the relation $\eta_{\rm tf} / \eta_{\rm hyd} \sim M_A^3$ for $M_A$ smaller than some 
$M_A^*$, 
which
increases with the decrease of the domain size. 
The theoretical limits for the validity of the classical weak turbulence 
indicated by the vertical lines in Figure~\ref{fig:ma_eta_tener_cs} are also shown 
in Figure~\ref{fig:ma_eta_tener_lx}.
Each color corresponds to a different domain size (following the same color 
scheme for each
set of simulations).
We also observe in the right panel of Figure~\ref{fig:ma_eta_tener_lx} that for 
simulations with $L_{\parallel} <$ 16L, the curve for $\tau_{\rm ener}/\tau_{\rm nl}$
deviates from a powerlaw for increasing values of $M_A^*$.

In the left panels of Figure~\ref{fig:ps_ts_1} we present
the 1D power spectrum of the total energy, $E(k_{\perp})$,
for simulations with different $M_A$. 
Inside these same panels we also show 
the ratio between the velocity and the magnetic field power spectra, 
$E_{\mathbf u}(k_{\perp}) / E_{\mathbf B}(k_{\perp})$.
On the right 
column we present the energy transfer spectrum, $T (k_{\perp})$, for the same models, 
normalized by $E_{\rm turb}/\tau_{\rm nl}$, which is  the expected value   of the energy transfer spectrum in the case of strong turbulence cascade.
The presence of a plato in the energy transfer spectrum 
is indicative of an inertial range, where the energy flux between scales is constant. 
The flatness of $T (k_{\perp})$ just after the injection scale 
$k_{\perp}L / 2\pi \approx 3$-$4$ is important to guarantee that with the current resolution, 
the energy transfer time
from this scale (and therefore the velocity correlation time) is shorter 
than the dissipation time.
If this were not the case, the measured $\eta_{tf}$ 
could be dominated by the 
numerical effective viscosity, as the velocity decorrelation time 
would be of the order 
of the dissipation time (see Equation~\ref{eq:diffusivity_general}).
The top panels  show simulations with the 
highest $M_S=0.32$, and domain size 16Lx1L.
The middle panels show models with the smallest $M_S=0.02$.
In both sets, the power spectrum of the model  with $M_A$ closer to unity shows a poor inertial range
($4 \lesssim k_{\perp} L / 2 \pi \lesssim 10$). 
For this wavenumber 
interval, the normalized transfer spectrum 
decays roughly ten percent for the model with $M_S=0.02$ (middle panel). 
As $M_A$ decreases, 
a gradual steepening in the spectrum is accompanied by the decrease on the energy transfer 
values at the injection scale, which is consistent with the increase of the energy transfer time.
The power law index seems to become steeper than $-2$ (the value predicted for the weak turbulence regime)
for the models with $M_S=0.02$ (middle panels).
The transfer spectra for the simulations with $M_S=0.02$ (middle panel) do not reveal 
a clear inertial range for 
the models with $M_A = 0.2$ and $M_A = 0.1$. Nonetheless, the microscopic (or numerical) dissipation does not seem to be
predominant at the perpendicular wavenumbers
just above the injection wavenumber $k_{\perp}L / 2\pi \approx 4$. We see that the transfer spectra have not
 been reduced 
significantly until the wavenumbers above $k_{\perp}L / 2\pi \approx 7$. The impact of the 
dissipation level close to the injection scale 
will be analyzed in Section 4.1 through the comparison of models with different perpedicular resolutions.
The bottom  panels  show models with $M_S=0.02$ and a 
domain size 1Lx1L, for which the validity of the classical weak turbulence has the 
theoretical lowest limit around $M_A = 0.5$. 
Below  this value,
the power spectrum does show the steepening seen in the model with extended domain (middle panel) with the decrease of $M_A$, and the transfer spectrum 
does not reduce substantially, which is consistent with a non increasing energy transfer time,  confirming  the previous analysis for these models (see the right panel of Figure~\ref{fig:ma_eta_tener_lx}).

The ratio between
the 
kinetic
and the magnetic 
power spectra 
shown in the left panels of Figure~\ref{fig:ps_ts_1} is 
approximately constant and close to unity, except for the largest wavenumbers,
inside the dissipation range, 
where numerical effects dominate.
When the energy spectrum is dominated by Alfv\'en modes, 
we should expect equipartition between the magnetic and kinetic energy spectra inside the inertial range, 
following the equality between these two amplitudes in each individual Alfv\'en mode.
We observe this equipartition even for our models with larger compressibility (models 16L-Ms0.32-A; 
see top left panel in Figure~\ref{fig:ps_ts_1}).

\begin{figure*}
\begin{tabular}{c c}
	\input{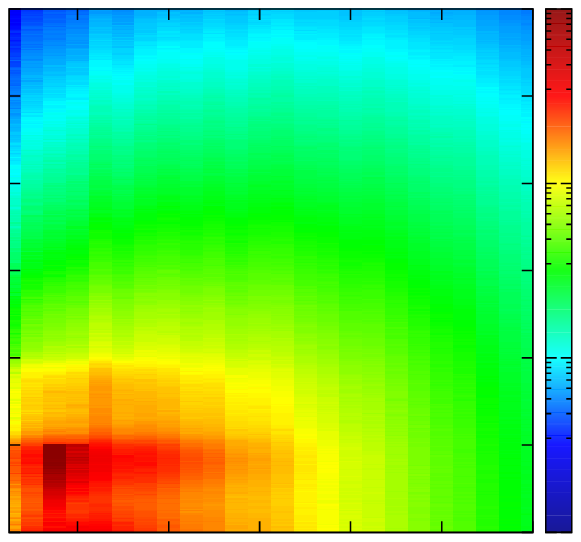} &
	\input{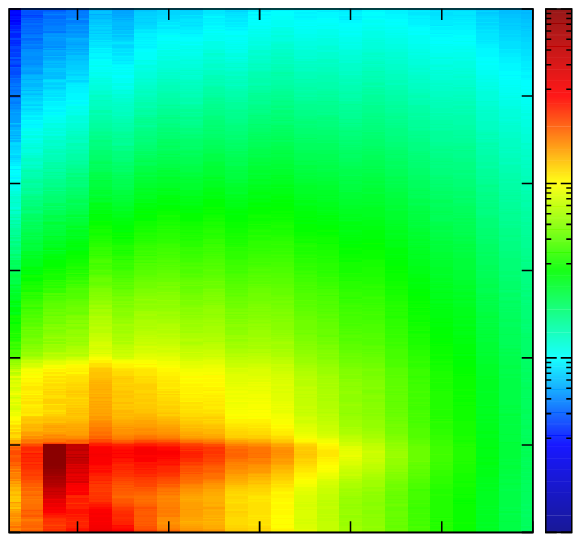} \\
	\input{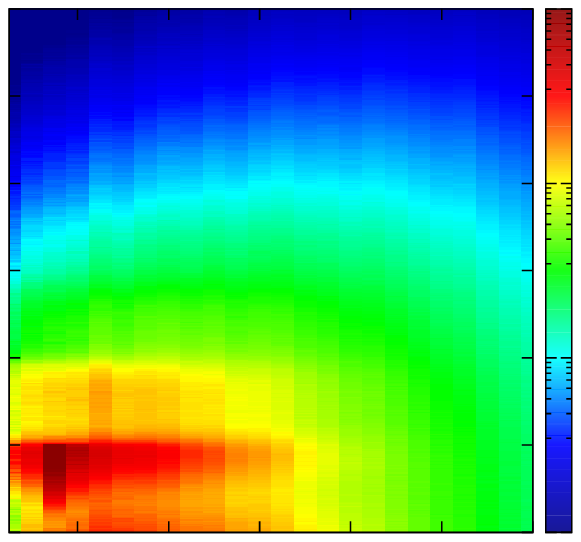} &
	\input{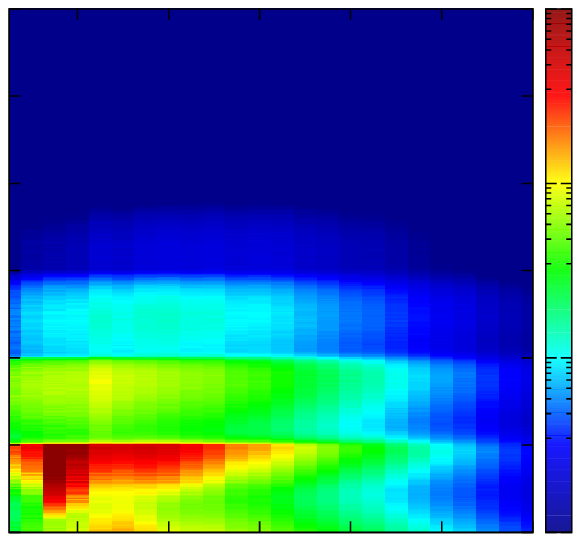} \\
	\input{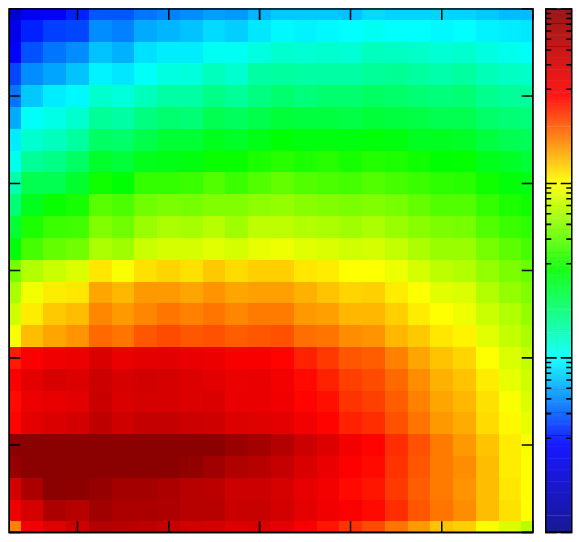} &
	\input{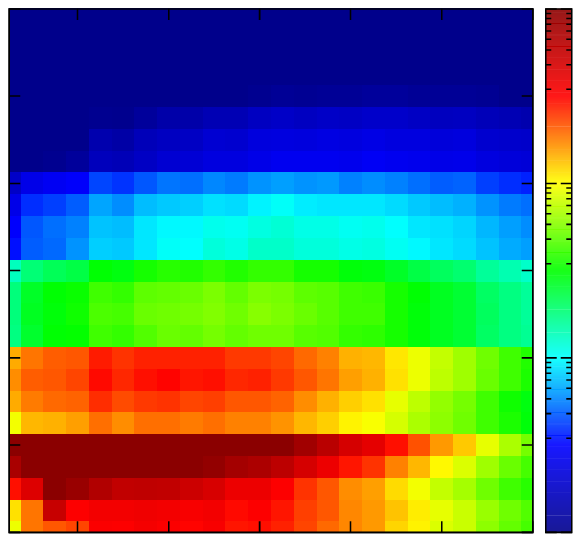} \\
\end{tabular}
	\caption{The 2D energy spectrum $E(k_{\parallel}, k_{\perp})$ distribution in the $(k_{\parallel}, k_{\perp})$ plane. 
	Each column corresponds to a different nominal Alfv\'{e}nic Mach number $M_{A,0} \equiv v_{0}/v_{A,0}$.
	Left column: $M_{A,0} = 0.4$. Right column: $M_{A,0} = 0.2$.
	Top row: domain size $L_{\parallel} =$ 16L and sonic Mach number $M_S=0.32$.
	Middle row: domain size $L_{\parallel} =$ 16L and sonic Mach number $M_S=0.02$.
	Bottom row: domain size $L_{\parallel} =$ 1L and sonic Mach number $M_S=0.02$.
        See Table~\ref{tab:runs_params_1} for the complete description of the simulations parameters. }
\label{fig:maps_ps_cs_LX}	
\end{figure*}

Figure~\ref{fig:maps_ps_cs_LX} compares the 2D energy spectrum for 
simulations from the same 
sets shown in Figure~\ref{fig:ps_ts_1}.
On the left column the simulations have $M_A=0.4$, and
on the right column $M_A=0.2$. Top and middle panels 
correspond to
simulations with fixed domain size,
$L_{\parallel}=16$L, but different sonic Mach numbers,
$M_S=0.32$ 
(top panels),
and $M_S=0.02$ (middle panels). The less compressible simulations 
($M_S=0.02$)
reveal two features which are not visible in the most compressible simulations ($M_S=0.32$): 
a suppression in the energy cascade in the parallel direction (vertical 
axis and steepening 
of the energy spectrum in the perpendicular direction (horizontal axis) 
when $M_A$ decreases. 
Both features are expected to emerge in the weak turbulence regime 
in the limit of incompressible MHD (Afv\'en waves turbulence).
The bottom panels of Figure~\ref{fig:maps_ps_cs_LX} 
correspond to simulations with the same 
compressibility as the middle panels 
($M_S=0.02$), but with a shorter domain size in the parallel direction $L_{\parallel}=1$L.
The two features described above for the simulations with larger domain size 
are clearly weaker in the simulations with shorter domain, which are out of the limit
given by Equation~\ref{eq:wwt_validity} (at least near the injection scale).

We note in Figure~\ref{fig:maps_ps_cs_LX} that the energy distributions in the 
parallel direction (vertical axis) present some peaks or ``steps''. This effect becomes stronger 
in the right 
panels, for which the mean magnetic field is stronger (smaller $M_A$). These peaks
appear around the parallel wavenumbers which are 
harmonics of 
the wavenumber where the forcing amplitude is maximum, $k_{\parallel} L/ 2 \pi = 4$ 
(see Table~\ref{tab:runs_params_1}).
We attribute this feature to the peaked
distribution
of the forced modes
on the parallel wavenumbers
for the A-forcing scheme. 
At the same time,
the wave-turbulence character becomes more pronounced when 
the mean magnetic field is stronger. Non-linear coupling between triads 
of Alfv\'en waves naturally generates the higher order harmonics.
The presence of these same harmonics 
in the energy spectrum
was also pointed out in 
\cite{ghosh_etal_2009}, 
for the turbulence produced by a spectrum of Alfv\'en waves containing only one discrete parallel frequency
(monochromatic) combined with quasi-2D MHD modes (nearly zero Alfv\'en frequency) which has
no broad enough spectrum around 
$k_{\parallel}=0$. 
In Figure~\ref{fig:maps_ps_F_ma} (discussed in the next subsection), it 
is shown that these steps are much less pronounced in the simulations with 
the Ab-forcing scheme, which has a broader and smoother distribution 
of  amplitudes in the parallel wavenumbers. In this case, the non-linear coupling of waves 
involving 
a broader spectrum 
is enough to ``fill'' the gaps around the peaks seen before. 
Therefore, while the presence of the steps 
in the parallel spectrum of the simulations 
with the A-forcing
makes the wave character of the turbulence visible, 
the resulting magnetic field diffusion and energy transfer time are not 
different from the models employing the Ab-forcing scheme, where 
the parallel spectrum is smooth.

\begin{figure*}
\begin{tabular}{c c}
	\input{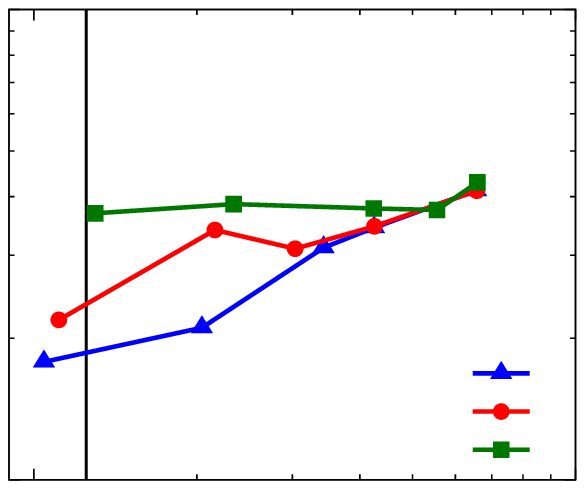} &
	\input{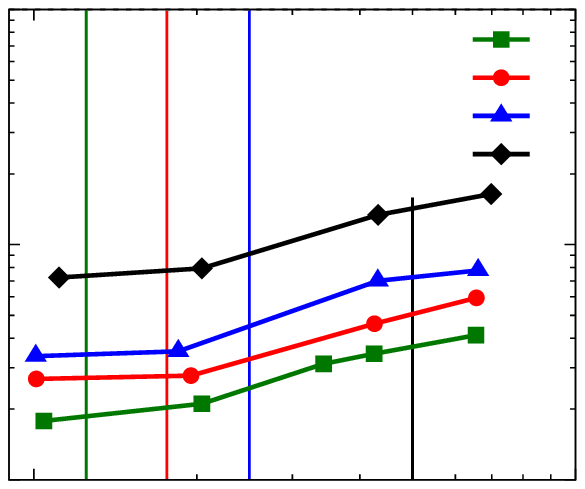} \\
\end{tabular}	
	\caption{Ratio between the rms value of the 2D component of the solenoidal velocity $\langle v_{\rm 2D,sol} \rangle$ and the total rms velocity $v_{\rm rms}$,
	as a function of the Alfv\'enic Mach number $M_A = v_{\rm rms}/\langle v_{A} \rangle$. 
	Left: simulations with different sonic Mach numbers $M_S = v_0/c_s$ are compared. 
	Right: simulations with the same $M_S = 0.02$, but different domain sizes are compared.
	Each point in the curves corresponds to a model in Table~\ref{tab:runs_params_1}. 
	For each domain size (corresponding to different  color curves on the right panel), 
	a vertical solid line indicates the
	lower limit of $M_A$ given by Eq.~\ref{eq:wwt_validity} at the injection scale $\ell$.}
\label{fig:ma_v2d_cs_LX}
\end{figure*}

Figure~\ref{fig:ma_v2d_cs_LX} shows, 
for the same set of simulations
presented in the Figures~\ref{fig:ma_eta_tener_cs} and \ref{fig:ma_eta_tener_lx}, 
the dependence  of the ratio between the amplitude 
of the 2D component of the solenoidal velocity field,
$\langle v_{\rm 2D,sol} \rangle$
and the total rms velocity $v_{\rm rms}$,
with $M_A$.
We calculate $\mathbf{v}_{\rm 2D,sol}$ from the Fourier components of the 
velocity field, $\mathbf{v}_\mathbf{k}$, by removing the vector components 
parallel to both $\mathbf{k}$ and $\mathbf{B}_0$ (that is, the vector components which are either potential 
or parallel to the mean magnetic field), and finally keeping only the modes for which $k_{\parallel}=0$).
The 2D velocity
components are not excited by the  A-forcing scheme, 
but they naturally develop in the system from  the wave interactions. 
In the weak turbulence theory, the 2D modes are required as one component in the three-wave resonant interactions 
(although they can exist only in finite sized domains). 
As these velocities do not bend the magnetic field, they can easily mix the field lines 
in the perpendicular direction, similar to hydrodynamical motions ($\eta_{\rm 2D} \sim \ell v_{2D}$), dominating 
the diffusion rate. 
Similar to Figure~\ref{fig:ma_eta_tener_cs}, 
the left panel of Figure~\ref{fig:ma_v2d_cs_LX} compares 
simulations with the same 
domain size 16Lx1L, but different compressibility ($M_S$). 
Analogous to the behavior of $\eta_{\rm tf}$
for $M_A > M_A^*$, the values of $\langle v_{\rm 2D,sol} \rangle / v_{\rm rms}$ converge to an 
approximate power law in $M_A$, 
and deviate from this 
trend for $M_A < M_A^*$
($M_A^* \approx 0.5, 0.4,$ and $0.2$ for 
$M_S = 0.32, 0.08,$ and $0.02$, respectively;
we see that 
these values of $M_A^*$ correspond to the value of $M_A$ 
above which the energy transfer time 
follows the dependence $\tau \propto M_A^{-2}$, as seen 
in the right panel of Figure~\ref{fig:ma_eta_tener_cs}).
Nonetheless, this analysis should be taken with 
caution because of the short range of values of the ratio $\langle v_{\rm 2D,sol} \rangle / v_{\rm rms}$
and the statistical uncertainties generated by the fluctuations in the curves. 
Following the trend of $\eta_{\rm tf}$, $\langle v_{\rm 2D,sol} \rangle /v_{\rm rms}$ also increases with $M_S$. 
The right panel of Figure~\ref{fig:ma_v2d_cs_LX} compares 
simulations with the same $M_S=0.02$, but different 
domain sizes. 
In this case all the sets of simulations show 
a similar qualitative behavior, i.e., 
$\langle v_{\rm 2D,sol} \rangle/v_{\rm rms}$ increases with $M_A$. 
The magnitude of each curve is inversely proportional to the square 
root of L$_{\parallel}$.
In summary, the magnitude of $\langle v_{\rm 2D,sol} \rangle$ follows 
qualitatively that of $\eta_{\rm tf}$, 
but the dependence of the diffusivity with $M_A$ is stronger for $\eta_{\rm tf}$. 
There is no direct evidence that the mix caused by the 2D motions dominates 
the turbulent diffusivity,  even for $M_A$ below $M_A^*$, for the 
simulations
presented in Figures~\ref{fig:ma_eta_tener_cs} and \ref{fig:ma_eta_tener_lx}.

\begin{figure*}
\begin{tabular}{c c}
	\input{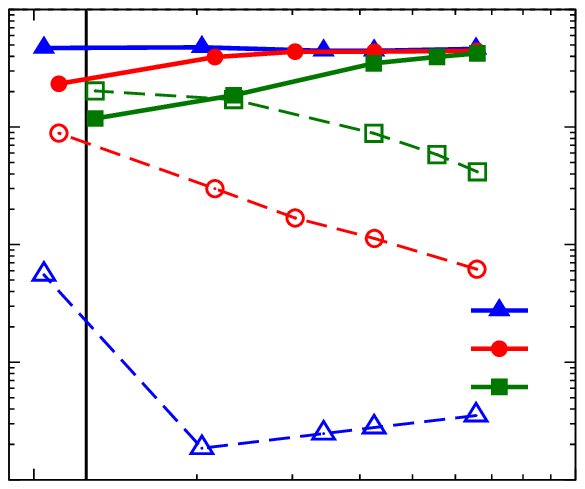} &
	\input{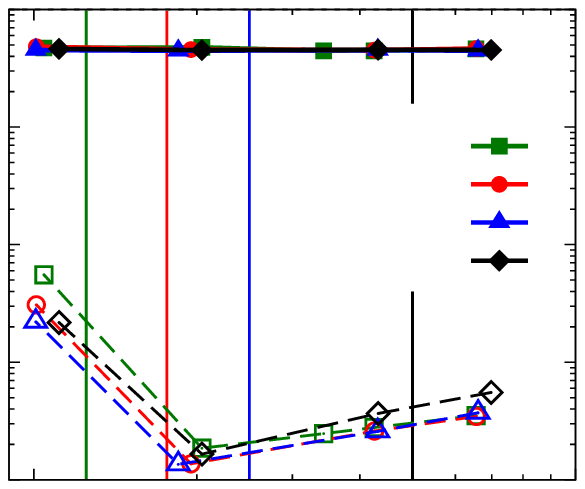} \\
\end{tabular}
	\caption{Ratio between the energy in each magnetosonic component of the 
	turbulence and the total turbulent energy $E_{\rm turb}$
	as a function of the Alfv\'enic Mach number $M_A = v_{\rm rms}/\langle v_{A} \rangle$. 
	The continuous lines are for the energy in the slow modes $E_{\rm slow}$, and the dashed lines are for the 
	energy in the fast modes $E_{\rm fast}$.
	Left: simulations with different sonic Mach numbers $M_S = v_0/c_s$ are compared. 
	Right: simulations with the same $M_S = 0.02$, but different domain sizes are compared.
	Each point corresponds to a model in Table~\ref{tab:runs_params_1}. 
	For each domain size (corresponding to different  color curves on the right panel), 
	a vertical solid line indicates the
	lower limit of $M_A$ given by Eq.~\ref{eq:wwt_validity} at the injection scale $\ell$.}
\label{fig:ma_ema_cs_LX}
\end{figure*}

In order to quantify the amount of turbulent energy in compressible modes for the simulations, 
for each wave vector in the Fourier space we performed 
the projection of the MHD variables onto the  
magnetosonic slow and fast eigenvectors.
Figure~\ref{fig:ma_ema_cs_LX} compares, 
for the same set of models presented in Figure~\ref{fig:ma_v2d_cs_LX}, 
the ratio between the turbulent energy in each magnetosonic mode (slow and fast) and the total turbulent energy, 
as a function of $M_A$.
The left panel compares models with different $M_S$ and fixed domain size 16Lx1L. 
For the most incompressible run set ($M_S = 0.02$), the 
relative energy in the slow modes is nearly constant with the different values of $M_A$, keeping close 
to $0.5$,
and increases slightly with the decrease of $M_A$. The relative energy in the fast modes is 
about two 
orders of magnitude below that in the slow modes, and decreases 
slowly with the decrease of $M_A$. Only for the simulation with the smallest value of 
$M_A$ there is a sudden increase in the relative energy of the fast modes.
The orther sets of simulations, with larger values of $M_S$,
show a trend for the decrease in the relative energy 
of the slow modes with the decrease of $M_A$, at the same time that the 
relative energy of the fast modes 
increases. For the most compressible simulation ($M_S=0.32$), for the smallest 
values of $M_A$ the energy in the fast modes overpass the energy of the slow modes.
The right panel of Figure~\ref{fig:ma_ema_cs_LX} compares models with fixed $M_S=0.02$ 
but different domain sizes.
All sets of simulations behave 
similarly to the model with the largest domain size shown in the left panel.

\subsection{Forcing effects}
\label{sec:results_forcing}

All the results described so far were
derived from simulations using the turbulence forcing anisotropically  distributed in the Fourier space (A-forcing).
In order to test the sensitivity of these results to
the forcing scheme, we have repeated the simulations with $M_S = 0.02$ and domain size 8Lx1L using two alternative forcings,
one isotropically distributed inside a spherical shell in the {\bf k}-space (I-forcing), and another where the amplitude of the 
modes inside a spherical shell are concentrated 
around $k_{\parallel}/k_{\perp} = 1$ using the modulation factor 
$\propto (k_{\parallel} k_{\perp} / k^2)$ 
(Ab-forcing). 
This last forcing scheme can be thought as intermediate 
between the extreme A- and I-forcing cases 
(such as described in Section~\ref{sec:setup}).

\begin{figure*}
\begin{tabular}{c c}
	\input{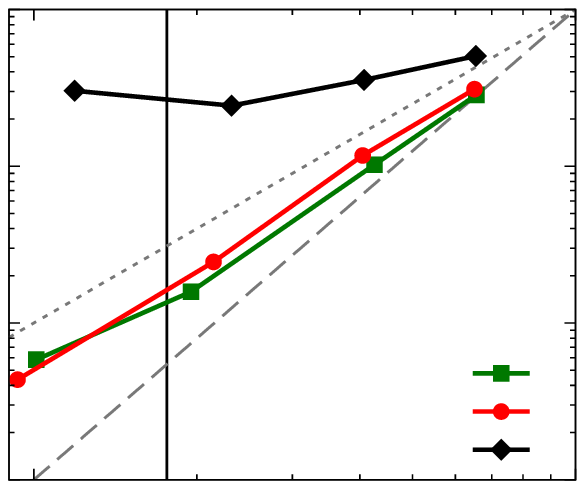} &
	\input{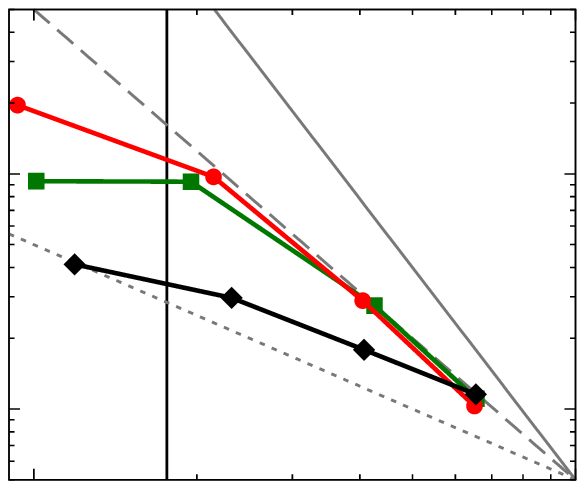} \\
\end{tabular}	
	\caption{Magnetic diffusion coefficients $\eta_{\rm tf}$ measured by the test-field method (left)
	and the energy transfer time $\tau_{\rm ener} \equiv E_{\rm turb} / T_{\rm turb}$ (right),
	as a function of the Alfv\'{e}nic Mach number $M_A = v_{\rm rms}/\langle v_{A} \rangle$. 
	Simulations with different turbulence forcing schemes are compared.
	All the runs have the same sonic Mach number $M_S = 0.02$ and domain size 8Lx1L.
	Each point corresponds to a model in Table~\ref{tab:runs_params_1}. 
	For the parameters used in these simulations the vertical solid line indicates the
	lower limit of $M_A$ given by Eq.~\ref{eq:wwt_validity} at the injection scale $\ell$.}
\label{fig:ma_eta_tener_f}
\end{figure*}

The left panel of Figure~\ref{fig:ma_eta_tener_f} shows that
for simulations with the I-forcing, 
$\eta_{\rm tf}$ is essentially independent of $M_A$ for $M_A < M_A^* = 0.2$.  Above $M_A^*$, 
this dependence follows a power law,
$\eta_{\rm tf} / \eta_{\rm hyd} \sim M_A^{-1}$, which is much weaker than 
the observed 
for the A-models, where the power low dependence is $\sim M_A^{-3}$.
The Ab-models behave similar to the A-models, with 
scaling of 
$\eta_{\rm tf} / \eta_{\rm hyd}$ between $\sim M_A^{-3}$ and $\sim M_A^{-2}$. 
In the right panel of Figure~\ref{fig:ma_eta_tener_f} we see
that the energy cascading times,
$\tau_{\rm ener}$, around the injection scale, $\ell$,
for the Ab-models are almost identical to those of  A-models, at least for $M_A > M_A^*$. 
In contrast, the increase of $\tau_{\rm ener}$ with the decrease of $M_A$ for the 
I-models is much slower than that for the A-models.

\begin{figure*}
\begin{tabular}{c c}
	\input{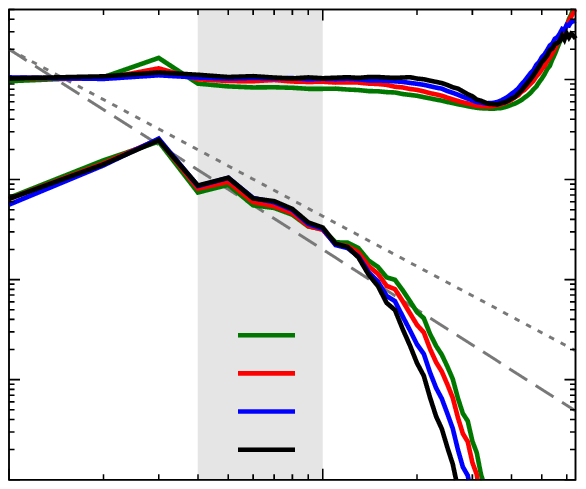} &
	\input{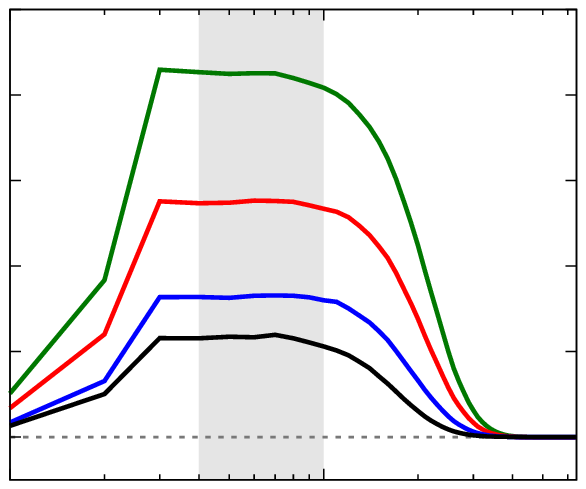} \\
	\input{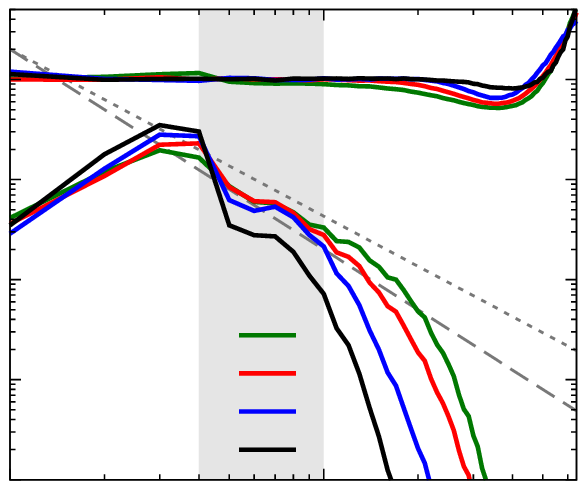} &
	\input{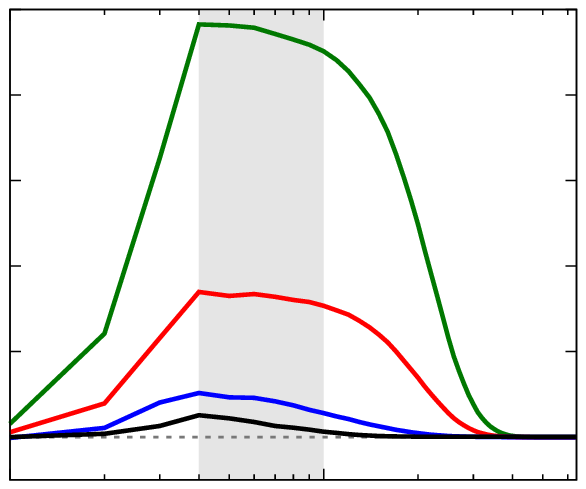} \\
\end{tabular}	
	\caption{The energy spectrum $E(k_{\perp})$ 
	(left column) and the energy transfer spectrum $T (k_{\perp})$ (right panel) for models with different turbulence forcing schemes. 
	The ratio between the velocity and the magnetic field power spectra for each model is shown inside the energy spectrum plots.
	In each panel, models from Table~\ref{tab:runs_params_1} with different Alfv\'enic Mach numbers $M_A$ are represented by curves with different colors.
	All the runs have the same sonic Mach number $M_S = 0.02$ and domain size 8Lx1L.
	Top: isotropic forcing scheme $I$; bottom: anisotropic forcing scheme $Ab$. 
	The power laws $\propto k_{\perp}^{-5/3}$ and $\propto k^{-2}$ are also depicted for comparison in the left panels.
	The gray area covers the wavenumbers range $4 < k_{\perp}L / 2 \pi < 10$ for which the transfer spectrum is approximately constant and close to its maximum value.}
\label{fig:ps_ts_2}
\end{figure*}

The turbulence energy spectrum $E(k_{\perp})$ for the simulations with I- and Ab-forcing 
are shown in the left column of Figure~\ref{fig:ps_ts_2}.
Their respective energy transfer spectrum, $T (k_{\perp})$,
are shown in the right column. We do not notice the steepening of the energy spectrum 
with the decrease of $M_A$ for the I-models as seen in the Ab-models and A-models (Figure~\ref{fig:ps_ts_1}).
Considering the analysis of both,
the energy cascading time (right panel of Figure~\ref{fig:ma_eta_tener_f}) and the 
energy transfer spectrum (right panel of Figure~\ref{fig:ps_ts_2}), there is a difference 
between the I-models and the $\mathbf{k}$-anisotropic forcing models A and Ab.
Notice that all the forcing schemes produce similar energy spectrum for $M_A = 0.8$, therefore, 
the difference is manifested only in the presence of strong magnetic fields.

\begin{figure*}
\begin{tabular}{c c}
	\input{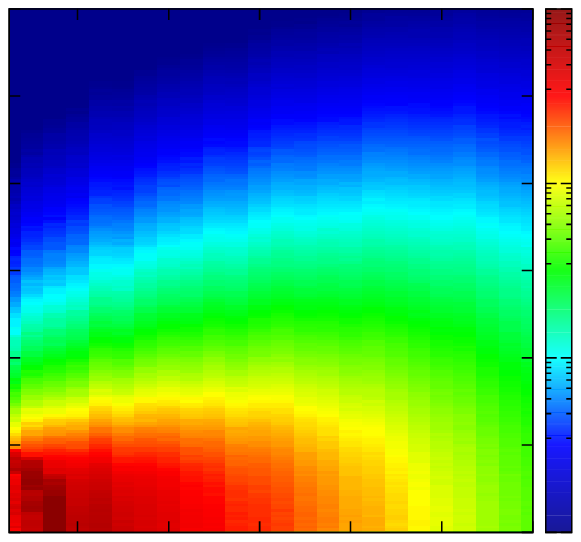} &
	\input{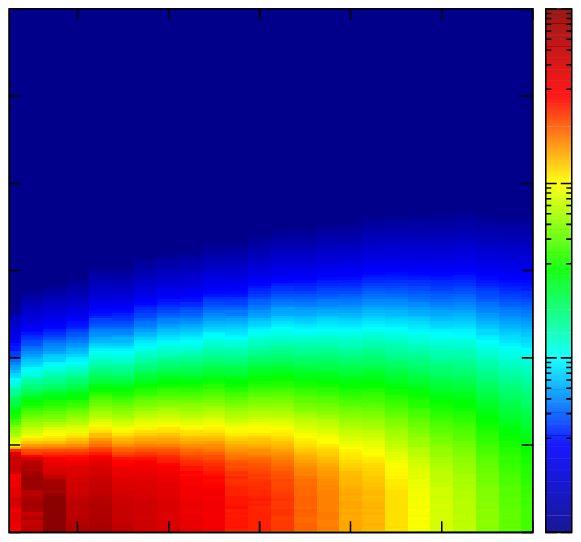} \\
	\input{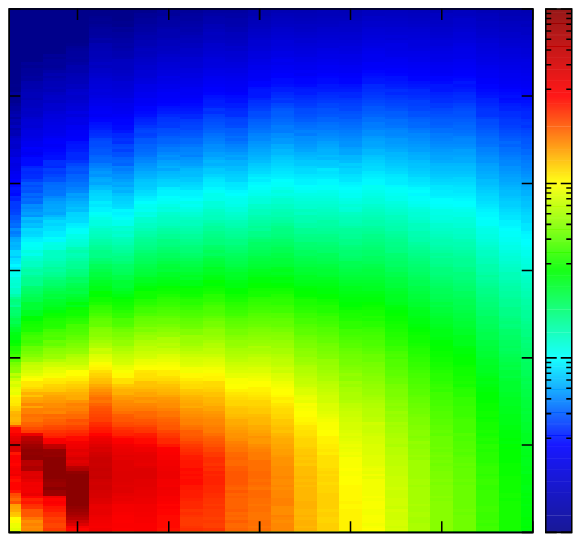} &
	\input{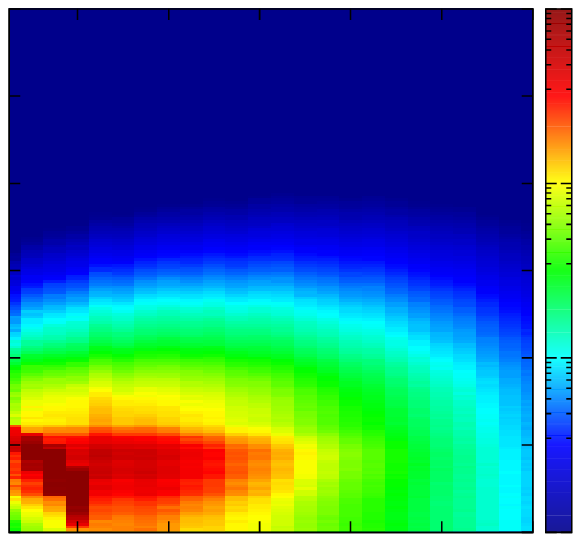} \\
	\input{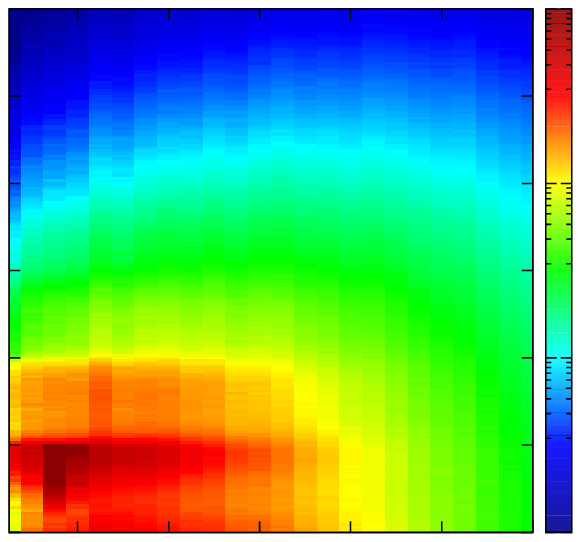} &
	\input{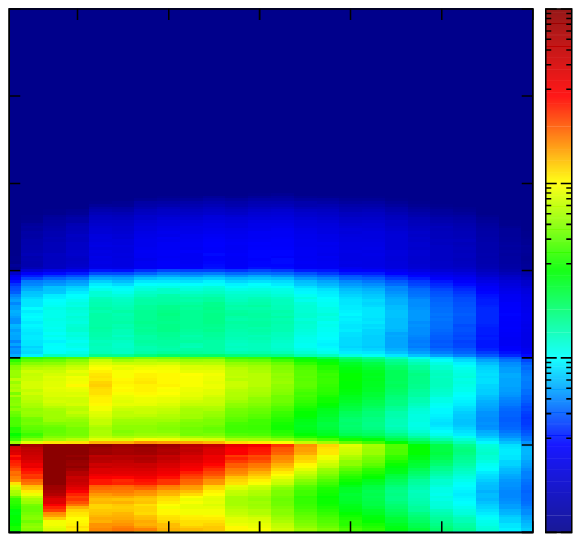} \\
\end{tabular}
	\caption{The 2D energy spectrum $E(k_{\parallel}, k_{\perp})$ distribution in the $(k_{\parallel}, k_{\perp})$ plane for models with different turbulence forcing schemes. 
	All the runs have the same sonic Mach number $M_S = 0.02$ and domain size 8Lx1L.
	Each column corresponds to a different nominal Alfv\'{e}nic Mach number $M_{A,0} \equiv v_{0}/v_{A,0}$.
	Left column: $M_{A,0} = 0.4$. Right column: $M_{A,0} = 0.2$.
	From top to bottom: I-forcing, Ab-forcing, and A-forcing. 
        See Table~\ref{tab:runs_params_1} for the complete description of the simulations parameters. }
\label{fig:maps_ps_F_ma}	
\end{figure*}

The 2D energy spectrum, $E(k_{\parallel}, k_{\perp})$ for 
simulations with the different forcing schemes are compared in 
Figure~\ref{fig:maps_ps_F_ma}. 
The top, middle and bottom panels correspond to the I, Ab, and A forcings, respectively.
Simulations in the left column have $M_A=0.4$, and in the right column $M_A=0.2$.
We note that substantial differences between models with different forcing become
evident for smaller values of $M_A$.
The 2D energy spectrum for values of $M_A$ closer to unity are almost indistinguishable (not shown). 
The $\mathbf{k}$-anisotropic models (middle and bottom panels) show steepening in the energy 
distribution in the perpendicular direction (horizontal axis), while 
this effect is not observed in the simulations with I forcing (top panels).
In all the cases we observe a sharp steepening of the spectrum in the parallel direction 
(vertical axis) for $M_A = 0.2$.

\begin{figure*}
\begin{tabular}{c c}
	\input{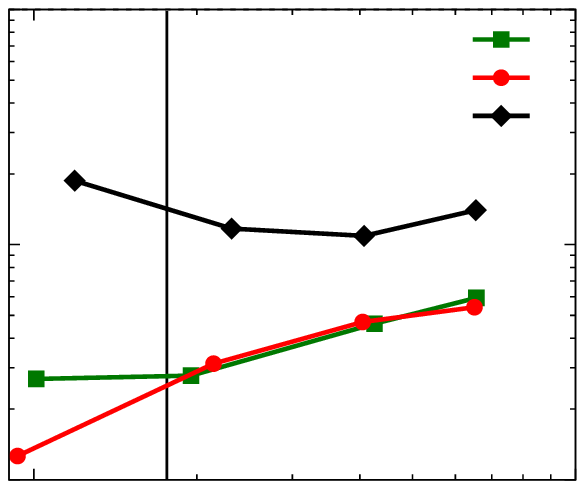} &
	\input{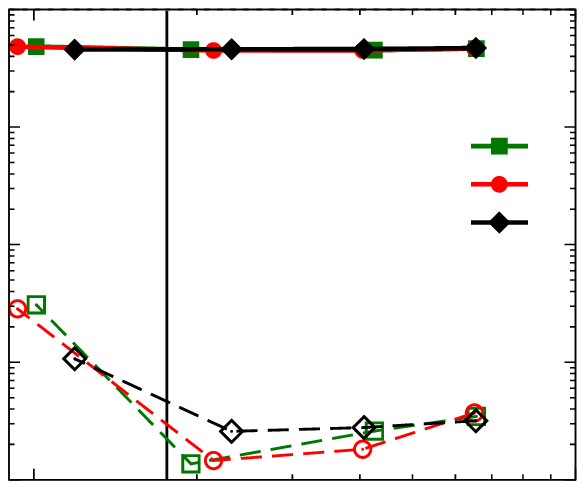} \\
\end{tabular}	
	\caption{Ratio between the rms value of the 2D component of the solenoidal velocity $\langle v_{\rm 2D,sol} \rangle$ and the total rms velocity $v_{\rm rms}$ (left),
	and the ratio between the energy in each magnetosonic component of the 
	turbulence and the total turbulent energy $E_{\rm turb}$ (right),
	as a function of the Alfv\'enic Mach number $M_A = v_{\rm rms}/\langle v_{A} \rangle$. 
	In the right panel, the continuous lines are for the energy in the slow modes $E_{\rm slow}$, and the dashed lines are for the 
	energy in the fast modes $E_{\rm fast}$.
 	Models with different turbulence forcing schemes are compared in each panel. 
	All the runs have the same sonic Mach number $M_S = 0.02$ and domain size 8Lx1L.
	Each point corresponds to a model in Table~\ref{tab:runs_params_1}. 
	For the parameters used in these simulations the vertical solid line indicates the
	lower limit of $M_A$ given by Eq.~\ref{eq:wwt_validity} at the injection scale $\ell$.}
\label{fig:ma_v2d_ema_F}
\end{figure*}

With exception of the I-cases, the dependence of the diffusivity $\eta_{\rm tf}$ with $M_A$ for all the 
simulations with different forcings seems to be closely 
related to the behavior of $\tau_{\rm ener}$ with $M_A$.
Indeed, the left panel of Figure~\ref{fig:ma_v2d_ema_F} shows that
the values of  
$\langle v_{\rm 2D,sol} \rangle / v_{\rm rms}$ 
for the I-models are much larger compared to the other forcing schemes. 
There is no surprise in this fact, as the 2D velocity modes are only forced 
in the I-models. These high 2D velocities can explain the diffusivity observed 
in the I-models, as $\eta_{\rm tf} / \eta_{\rm hyd} \sim \langle v_{2D,sol} / v_{\rm rms} \rangle$. 

Finally, 
the right panel of Figure~\ref{fig:ma_v2d_ema_F} compares the relative turbulent energies in 
the magnetosonic slow and fast modes for the sets of simulations
with the several forcings. No appreciable difference is observed between them.

\subsection{Resolution effects and convergence}

\begin{figure*}
\begin{tabular}{c c}
	\input{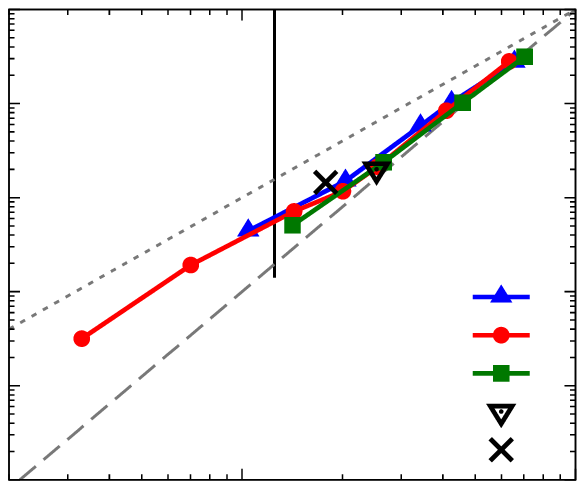} &
	\input{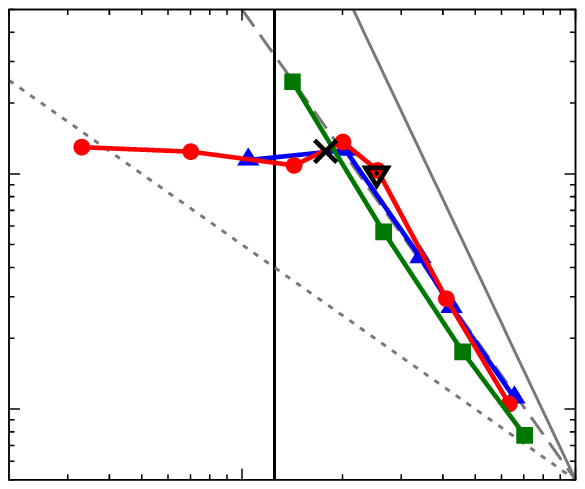} \\
	\input{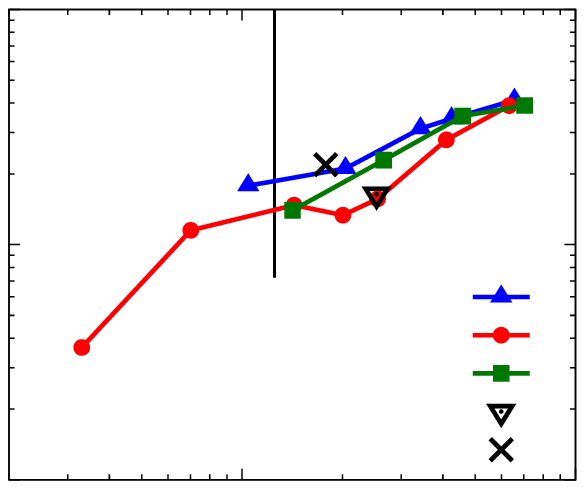} &
	\input{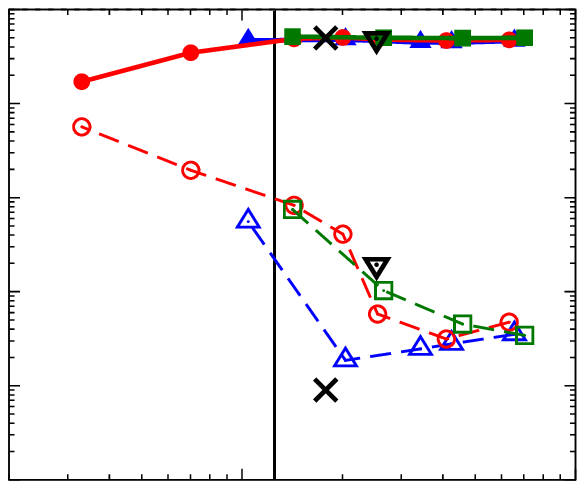} \\
\end{tabular}	
	\caption{
	Comparison between models 
        with different 
        numerical resolution
	(16L-Ms0.02-A, 16L-Ms0.02-low-A, 16L-Ms0.02-hi-A). In addition, models  16L-Ms0.02-low-A are compared 
	with identical counterpart models differing  only by the length of the domain perpendicular to the uniform 
	magnetic field
	(16x2L-Ms0.02-low-A), or by the replacement  of the hyper-viscosity and hyper-resistivity 
	by the standard viscosity and resistivity (16L-Ms0.02-low-A-diff2).
 	The  plots  show different  quantities versus the Alfv\'enic Mach number $M_A = v_{\rm rms}/\langle v_{A} \rangle$. 
	Top left: magnetic diffusion coefficients $\eta_{\rm tf}$ measured by the test-field method.
	Top right: energy transfer time $\tau_{\rm ener} \equiv E_{\rm turb} / T_{\rm turb}$.
	Bottom left: ratio between the rms value of the 2D component of the solenoidal velocity 
	$\langle v_{\rm 2D,sol} \rangle$ and the total rms velocity $v_{\rm rms}$.
	Bottom right: ratio between the energy in each magnetosonic component of the 
	turbulence and the total turbulent energy $E_{\rm turb}$.
	The continuous lines are for the energy in the slow modes $E_{\rm slow}$, and the dashed lines are for the 
	energy in the fast modes $E_{\rm fast}$.
	All the runs have the same sonic Mach number $M_S = 0.02$ and domain size 16Lx1L.
	Each point corresponds to a model in Table~\ref{tab:runs_params_1}. 
	For the parameters used in these simulations the vertical solid line indicates the
	lower limit of $M_A$ given by Eq.~\ref{eq:wwt_validity} at the injection scale $\ell$.}
\label{fig:convergence} 
\end{figure*}

Figure~\ref{fig:convergence} compares the model 16L-Ms0.02-A having resolution 2048x128$^2$ with 
the models 16L-Ms0.02-low-A and 16L-Ms0.02-hi-A, with identical parameters but 
resolutions 
1024x64$^2$
and 
2048x256$^2$, respectively. In order to make the higher resolution run feasible 
with the computational power we have available,
the models 
16L-Ms0.02-hi-A 
have twice the resolution only in the 
perpendicular direction to the uniform magnetic field. 
In the regime of weak turbulence, we expect  (based on the theory) 
the non-linear energy transfer (turbulence cascade)
to be more important in 
this
direction, 
at least 
close to the injection scale. 

Moreover, in Figure~\ref{fig:convergence}, we compare the low resolution models
16L-Ms-0.02-low-A with other two models, 16Lx2L-Ms-0.02-low-A and 16L-Ms0.02-low-A-diff2.
The first of these models has double domain size in the direction
perpendicular to the mean magnetic field $L_{\perp}$, that is, 
it has the ratio $L_{\perp} / \ell_{\perp}$ increased by a factor of two
compared to the models 16L-Ms0.02-low-A.
The other models, 16-Ms-0.02-low-A-diff2, have the hyper-viscosity and 
hyper-diffusivity replaced by the standard viscosity and resistivity, 
respectively
(see the description of these microscopic diffusive terms in Sec.~\ref{sec:numerical_models}).

The minimum $M_A$ below which 
the 3D MHD turbulence becomes ``enslaved'' to the 2D modes 
is theoretically predicted to depend on the ratio $L_{\perp} / \ell_{\perp}$ 
(\citealt{nazarenko2007}; see Eq.~\ref{eq:2d_slaving}).
This minimum $M_A$ is indeed below the value delimiting the validity of the weak turbulence theory, 
which we have marked in our plots. Therefore, based on theory only, we should not expect differences 
in the turbulence regime between the models 
16L-Ms-0.02-low-A and 16Lx2L-Ms-0.02-low-A. Nonetheless, because these 
theoretical expressions give only order of magnitude estimates, we 
present this convergence test for the perpendicular size of the domain, in order 
to check whether 
or not
the length used in all the previous analyses 
influences our results.

The transport rate of large scale fields via reconnection diffusion (RD) 
is expected to be dominated by the statistics of the
larger scale motions of the turbulence. Since 
the Reynolds and the magnetic Reynolds numbers 
of
the turbulence flow are much larger than unity, 
the microphysics describing 
the diffusion mechanism (as for example ambipolar diffusion, 
ohmic resistivity, anomalous resistivity, or even ``numerical 
diffusivity'')
should not have impact on the magnetic field diffusion coefficient
(see Eq.~\ref{eq:diffusivity_general}). 
However, if the diffusion 
mechanism can somehow change the statistics of the turbulence 
close to the injection scale, then it can influence 
the RD process. The comparison between models 
16L-Ms-0.02-low-A and 16L-Ms0.02-low-A-diff2 aims
to check if the hyper-viscosity and hyper-resistivity, used 
in all the other models in this work, could have some 
effect on the magnetic diffusivity coefficients.

The upper-left panel 
Fig.~\ref{fig:convergence}
shows the diffusion coefficients $\eta_{\rm tf}$ as a function of 
$M_A$ 
(see also left panel of Figure~\ref{fig:ma_eta_tener_cs}). 
The scaling laws are 
similar for the different resolutions. 
The change in the domains size $L_{\perp}$ also does not 
results in important changes
in $\eta_{\rm tf}$. 
Finally, 
the results 
of 
the models using 
standard 
viscosity and hyperdifusivity (16L-Ms0.02-low-A-diff2) 
show no noticieable differences.

The low resolution models 16L-Ms-0.02-low-A contain runs with values of $M_A$ 
below the line indicating the
lower limit of $M_A$ given by Eq.~\ref{eq:wwt_validity} at the injection scale $\ell$. 
This extension shows more clearly the asymptotic change of 
the power-law dependence with $M_A$ 
from  $\eta_{\rm tf} / \eta_{\rm hyd} \sim M_A^3$
to $\eta_{\rm tf} / \eta_{\rm hyd} \sim M_A^2$,
as seen in the models with standard resolution of this work 
(see Figures~\ref{fig:ma_eta_tener_cs} and ~\ref{fig:ma_eta_tener_lx}). 

The
top right panel 
of Fig.~\ref{fig:convergence}
compares the  normalized 
energy transfer time from the injection scale, $\tau_{\rm ener}$,
for these simulation sets (see also right panel of Figure~\ref{fig:ma_eta_tener_cs}).
We observe that the curve for the set of simulations 
with lower resolution is slightly steeper, 
with the power law index in $M_A$ 
between $-2$ and $-3$ 
(for $M_A$ larger than some $M_A^*$, below which the curves become almost flat). 
This difference of power laws in $\tau_{\rm ener}$ 
probably reflects the almost imperceptible steeper power law 
in $\eta_{\rm tf}$ for the lower resolution model (top left panel).
The break in the curve showing the change in the dependence of $\tau_{\rm ener}$ with $M_A$
(at $M_A^*$) is not evident in the
runs set employing standard viscosity and resistivity.

Some small differences between 
the compared models 
are
also visible 
in the lower left panel of
Figure~\ref{fig:convergence},
which 
depicts
the relation between
$\langle v_{\rm 2D,sol} \rangle / v_{\rm rms}$ 
and $M_A$ 
(see also Figure~\ref{fig:ma_v2d_cs_LX}). 
The higher resolution models have relatively more energy 
in the 2D modes
(although the maximum difference is still less than a factor of 2).
Interestingly, when comparing
the energies in the magnetosonic modes for the two resolutions 
(bottom right panel of Figure~\ref{fig:convergence}), 
we notice 
that the lower resolution runs show the same increase in the energy of the fast modes 
seen in the medium resolution ones for the smallest values of $M_A$. 
Nonetheless, this increase happens for 
higher values of $M_A$ in the lower resolution
run, contrary to the reduction in the 2D solenoidal velocity fields.

All the comparisons in Figure~\ref{fig:convergence} 
suggest that the results for the simulations 
with the standard resolution (models 16L-Ms0.02-A, 
having the same cells-size
employed in all the simulations presented 
before in this 
section) may be close to convergence 
with respect to the numerical resolution, at least for 
$\eta_{\rm tf}$ in the interval of $M_A$ considered. 
Also, the results discussed in the previous sections seem 
not to be sensitive to changes in the domain size in the direction perpendicular 
to the mean magnetic field or to the use of the microscopic hyper-diffusivities 
instead of the standard 
dissipative terms.

We should also mention that the microscopic resistivity employed 
in the induction equation of the test-fields (see Appendix~\ref{ap:test-field}) is the standard resistivity
instead of the hyper-resistivity employed in the main induction equation.
The test-field magnetic diffusivity was kept fixed for all our simulations
$\eta_1 = 1.2 \times 10^{-2} \; \ell_{\perp} v_0$ (a value relatively high in order 
to keep the test-fields smooth for longer times). 
As discussed in the beginning of this section, we do not expect the microscopic terms 
to influence the diffusion rate of the large scale magnetic field.
Nonetheless, in order to verify if this 
inconsistency between the microscopic resistive 
terms
could 
influence the measurement of the turbulent magnetic diffusivity $\eta_{\rm tf}$, 
we repeated the run 16L-Ms0.02-low-A, $M_A = 0.4$, using hyper-resisitivity for the test-fields. 
The result for  $\eta_{\rm tf}$ (not shown here) revealed no difference.

\section{Discussion}

The diffusion coefficient predicted by the reconnection diffusion (RD) theory 
in the sub-Alfv\'{e}nic regime has been  derived 
for the scenario of purely Alfv\'{e}nic turbulence (i.e., the incompressible limit) in 
the weak regime. In addition, 
the scaling laws of the inertial range are
assumed to be valid also at the injection scale, as the largest scale motions 
are responsible for the diffusion rate.
The analysis of our simulations does not evince,
entirely,
the development of 
weak turbulence. 
The resolution available does not allow the analysis of the power law index 
of the turbulence spectrum, 
usually employed to characterize the turbulence.
However, our estimate of the energy transfer rate at the injection scales 
shows a dependence $\propto M_A^{-\alpha}$ converging to $\alpha = 2-3$, which is much 
stronger than what is expected by the weak turbulence theory (WTT) at the inertial 
range, $\alpha = 1$.
It should be pointed out that, 
to our knowledge there are no results in the literature showing a simulation 
of forced MHD turbulence 
reproducing clearly the results of the WTT. 
However, \cite{meyrand_etal_2014} (see also \citealt{meyrand_etal_2018}) 
identified the WT regime at scales well 
below the outer scale in a simulation of decaying turbulence. Employing a spectral MHD code, 
\cite{perez_boldyrev_2008} performed simulations of forced reduced MHD turbulence and observed a 
perpendicular power spectrum consistent with the WTT
when the injection is performed in a broad range of parallel wave-numbers, 
although
far from the scenario of isotropic injection.
From the observational side there is no clear evidence of the 
MHD
WT until the date. 
The only observation that could be consistent with the WTT is the power spectrum of one 
of the magnetic field components in the Jupiter's magnetosphere 
(see \citealt{saur_etal_2002}).

Despite of the lack of numerical simulations of forced MHD turbulence demonstrating 
the development of the WT for the cascade immediately from 
the 
isotropic
injection scale and below, there are several studies (for example 
\citealt{alexakis2011, alexakis2012, bigot_galtier_2011}) 
exploring forced turbulence in the sub-Alfv\'{e}nic regime, with isotropic forcing 
(i.e., strong turbulence is not imposed by anisotropic forcing).
These studies show the effects of limited box sizes,
such as the development of a 2D dominated cascade, inverse cascade, and the influence 
of different types of forcing (selection of modes in the \textbf{k}-space), but none of them 
was able to reproduce the WT power spectrum.

In summary, although the WTT for Alfv\'{e}n waves is relatively well founded, 
it seems challenging to produce simulations of forced MHD turbulence in this regime in 
order to test theories based on the WT scenario (as for example the predictions related 
to turbulent reconnection, RD, and cosmic ray diffusion). Besides, a robust  observational 
evidence of the MHD WT in nature is still missing.

The simulations presented in this study are not incompressible. However, we kept 
the induced turbulence subsonic and, although the dominance of the Alfv\'{e}n waves  
in the energy spectrum, some level of magneto-sonic waves is seen. 
The compressible code employed (the \textsc{Pencil Code}) has the advantage to have implemented 
the test-field method tested in a large number of studies 
to measure with precision the turbulent diffusion coefficient of 
the magnetic field. Although the scenario of the RD theory in purely incompressible,
some degree of compressibility is expected in the turbulence in astrophysical environments 
(as for example, in star forming regions of molecular clouds).
There is still no theory of weak turbulence for a compressible MHD gas. 
The cascade of fast modes is 
weak~\footnote{Weak here means that the interactions decrease with the amplitude of the waves, 
i.e. $\sim v_{\ell}/\sqrt{c_s^2 + V_A^2}$, where $c_s$ is the sound velocity.  The corresponding spectrum
corresponds to $\sim k^{-3/2}$. The simulations in \cite{kowal_lazarian_2010} show that the power spectrum 
of fast modes can be shock-like, i.e. $\sim k^{-2}$. The latter result still requires further studies with
other codes.} 
at least in subsonic turbulence 
(see \citealt{cho_lazarian_2002}). Slow modes only cascade fast in the presence 
of strong Alfv\'enic turbulence.
In the future, it is important to confirm our conclusions for the validity of the RD 
diffusivity prediction in the asymptotic incompressible limit using for example a 
spectral MHD code.

In view of all the points raised above, we cannot discard that the turbulence regime developed 
in our simulations is intermediary between the asymptotic weak and strong limits.
This issue can only be investigated by further analysis of simulations with improved 
resolution. At present, our simulations do not have a clearly identifiable inertial range, 
although our analysis of the energy transfer spectrum indicates
that the numerical dissipation at the injection scales 
(where the magnetic diffusivity is produced) operates at a rate 
lower 
than the cascade/decorrelation rate.

It is therefore surprising that, despite the turbulence statistics of our simulations cannot be 
directly related to the scenario of the RD theory,
we still obtain results close to the predictions
(at least when we consider the setups
more suitable for the development of the WT).

It is also clear that 
compressibility modifies the dependence of the diffusion with $M_A$, 
making the diffusion rate closer to the dependence expected when turbulence 
is in the strong regime 
(i.e., when the critical balance $\ell_{\parallel}/v_a \sim \ell/\delta u_{\ell}$ is satisfied), 
and we have $\tau_{\rm casc} \sim (\ell / \delta u_{\ell})$ in eq. (\ref{eq:diffusivity_general}). 
In this situation the RD prediction would modify eq. (\ref{eq:eta_rd_prediction}) to 
$\eta_{\rm RD} \sim D_{\perp} \sim \ell \delta u_{\ell} \min \left( 1, M_A^{2} \right)$.
However, because the amplitude of the 2D modes of the velocity field 
increases with the sonic Mach number, it was not possible to disentangle 
their effects.
It remains to explore the turbulent diffusivity in the presence of 
turbulence more strongly compressible, which is more realistic, e.g.,  for the 
interstellar medium. It could be performed by implementing the test-field method in 
a compressible, shock-capturing MHD code.

Two different forcing schemes for the anisotropic amplitude distribution produced rather similar results.
Therefore, the results are apparently not very sensitive 
to the details of the ``\textbf{k}-anisotropic'' forcing, provided that
the amplitude of the forced modes decreases to zero when the parallel wavenumber goes to zero.
One way to strengthen this conclusion in the future would be to repeat this study using 
a different scheme for driving the turbulence.
One particularly interesting scheme consists in inducing random distributions of finite eddies directly in real space. 
This method was introduced in \cite{kowal_etal_2012}, 
where it was shown to produce the 
same turbulent reconnection rates as numerical simulations with forcing controled 
by their Fourier components 
(as is done in the present work).

Our results show that
the power in the two-dimensional modes, which can eventually 
dominate the turbulence cascade or dominate the diffusion of the magnetic field,
are only controlled when using anisotropic (in \textbf{k}-space) forcing schemes with 
domains sufficiently elongated in the direction parallel to the mean magnetic field.
These results are in agreement with the theoretical limits presented in \cite{nazarenko2007}. 

A systematic study of the turbulent diffusivity for sub-Alfv\'{e}nic turbulence with a mean magnetic 
field has been carried out in \citet[\citetalias{karak_etal_2014}]{karak_etal_2014}. This work also employed the \textsc{Pencil Code} 
with the test-field method to measure the diffusion coefficients. One of the setups studied in 
\citetalias{karak_etal_2014} uses non-helical random forcing, exciting modes isotropically in the \textbf{k}-space. With 
a square domain and values extremely low of $M_A$, the turbulent regime in their simulations could, 
at least theoretically, be in the discrete WT regime and enslaved to the 2D cascade (\citealt{nazarenko2007}; 
see Eqs.~\ref{eq:wwt_validity}-\ref{eq:2d_slaving}).
The dependence of the diffusion coefficient with $M_A$ observed in \citetalias{karak_etal_2014} (described as a quenching) 
is close to $M_A^{-1}$ for $0.05 \lesssim M_A \lesssim 1$, i.e., similar to that  we observed  in our simulations 
with isotropic forcing.
For the smallest values of $M_A$, the 
dependence with of $\eta_{\mathrm{turb}}$ with $M_A$ disappears, which is expected when the diffusion is 
dominated by the 2D velocity modes.
We can interpret the results of these authors as effects of the simulation setup, 
specially the domain size.

In the context of numerical simulations of turbulent reconnection, 
the increase of the reconnection rate 
could also arise
as an artifact of the limited domain size.
According to the discussion in the previous paragraphs,  in order to avoid this effect,  the domain size in the direction parallel to the large-scale magnetic field has to be large enough in these numerical studies.

The present study also calls for a re-evaluation of the transport coefficients in the 
large scale dynamo context 
(for example for 
helical turbulence), 
taking into account all role of compressibility and domain size of the simulation.
It is not clear if the same effects also affect the local simulations aimed to extract the mean field 
coefficients in the convective layer of the Sun 
or in accretion disks.  

The present study supports the predictions of the RD theory at least in the incompressible limit, 
but it also points to an increase of the diffusivity rate due to the 
compressibility of the turbulence, alleviating the strong suppression caused by the mean magnetic field. 
The RD mechanism has been proposed (\citealt{lazarian2005}) and tested 
successfully (at least qualitatively) to solve problems related to star formation: 
the magnetic flux problem (\citealt{santos-lima_etal_2010, leao_etal_2013}) and the magnetic braking 
catastrophe (\citealt{santos-lima_etal_2012, santos-lima_etal_2013, gonzalez-casanova_etal_2016}). 
\cite{lazarian_etal_2012} conclude from observations of molecular cloud cores 
 that the mass-to-flux ratio of 
super-critical cores, compared to their envelopes, are more consistent with the transport of magnetic flux 
via RD than ambipolar diffusion (AD).
More recent works on simulations of star formation processes, encompassing scales from molecular cloud clumps down to protostellar disks through 
the use of adaptive mesh, try to disentangle the role of other mechanisms that could solve these problems as well, 
like the 
misalignment of the angular momentum of the protostellar 
disk and the mean magnetic field 
(\citealt{joos_etal_2012, joos_etal_2013}),
the interchange instability, 
and the ambipolar diffusion
(see, for example, \citealt{hennebelle_inutsuka_2019} and references therein). 
These studies in general do not quantify the turbulent magnetic flux transport 
(due to the inherent difficult to perform this measurement), and it is far from clear if their resolution is able to represent the RD in these global simulations.
For example, \cite{lam_etal_2019} 
concluded that only 
a combination of turbulence and AD 
(each one dominating in different phases and/or regions)
can allow the formation of rotationally supported protostellar disks which persist 
for sufficiently long times.
We remark, however, that
the combination ``turbulent ambipolar diffusion''
is an inconsistent concept:
 when there is turbulence, this means that the AD is subdominant; 
if AD is very strong, there is no turbulence (see a more complete discussion in the recent 
review by \citealt{lazarian_etal_2020}).
The setting and interpretation of such 
global simulations depend on a full understanding of the RD mechanism driven 
by turbulence and the numerical simulation setup effects that can interfere
in the RD diffusivity. 
The present study provides a contribution in this direction. 
Further study of the RD in the presence of gravity are required as the properties of turbulence can be
modified by gravity. In fact, \cite{santos-lima_etal_2010} found some evidences 
that the transport of magnetic flux via RD increases with the intensity of the gravitational potential.

\section{Summary and Conclusions}

In this work we tested numerically the dependence of the magnetic diffusion coefficient 
provided by  reconnection induced by turbulence (Reconnection Diffusion, RD) 
with the Alfv\'{e}nic Mach number of the turbulence.
In all our 3D MHD simulations we imposed an  initially uniform magnetic field and focused on the sub-Alfv\'{e}nic regime. 
The turbulence is forced isotropically in the space.
Although  we  envision applications of the results, e.g.,  to  studies of the role of the RD during star-formation, inside molecular clouds 
where turbulence is expected to be trans-sonic or supersonic 
(\citealt{santos-lima_etal_2010, santos-lima_etal_2012, santos-lima_etal_2013, leao_etal_2013, gonzalez-casanova_etal_2016}), in this 
study we have focused only in the subsonic case.
The motivation is the direct comparison with the 
current RD theory, built on the scaling laws provided by the Alfv\'{e}nic turbulence in 
the weak regime (for sub-Alfv\'{e}nic isotropic injection).
We employed the \textsc{Pencil Code} with the Test-Field method to extract the average 
diffusion coefficient from the simulations.

The RD theory assumes that the inertial range scale laws of the weak turbulence theory can be extended to the 
injection scales, leading to a diffusion coefficient $\eta$ proportional to the hydrodynamical value multiplied by  the 
third power of $M_A$, when $M_A < 1$. 
We found no clear evidence of the development of the weak turbulence regime in our numerical simulations. In particular, 
the cascading time $\propto M_A^{-1}$ from the weak turbulence theory at the injection scale is not observed in any of our model sets.
Due to limited resolution and the fast increase of the cascading time with the increase of the magnetic field intensity, 
our simulations do not show appreciable inertial range to allow a robust determination  of the power law index of the power spectrum.
Nonetheless, the diffusion coefficients we obtain seem to be consistent with the RD prediction $\eta \propto M_A^3$ when the 
domain size parallel 
to the uniform magnetic field is large enough to avoid the finite box size effects 
(see \citealt{nazarenko2007} in the framework of reduced MHD) and the sonic Mach number small enough
($M_S \lesssim 0.02$). 
For smaller boxes and bigger values of $M_S$, we observed a dependence of $\eta$ more consistent with 
$M_A^2$, which 
could be 
the expected dependency in the strong cascading regime.

In the future, we will investigate both the incompressible limit to confirm the validity of the RD theory in a larger interval of $M_A$, and 
also the diffusivity provided by super-sonic turbulence which is more realistic for star-forming 
environments. 
At the same time, more numerical investigation is necessary for the weak turbulence regime, 
as it is not yet clearly reproducible in direct simulations of 
forced turbulence.

Due to the omnipresence of MHD turbulence in astrophysics, 
the proper understanding of the turbulent diffusivity is of fundamental 
importance not only in the context of star-formation, but it has also 
important consequences for the more general reconnection problem, 
for the large scale turbulent dynamo operating in all scales 
(stars, accretion disks, galaxies; see for example 
\citealt{xu_lazarian_2020} for
a recent study on the nonlinear turbulent dynamo in a gravitationally collapsing system
which accounts for the RD effects),
and for the propagation and acceleration of 
cosmic rays in astrophysical media.

\section*{Acknowledgements}
\addcontentsline{toc}{section}{Acknowledgements}

The authors thank A. Brandenburg, J. Cho, 
G. Eyink, B. Raphaldini, 
M. Rheinhardt,
M. V. del Valle,
and S. Xu for very useful discussions and comments.
The authors also thank the anonymous referee who helped to improve this work with her/his comments.
RSL acknowledges partial support from a grant of the Brazilian Agency FAPESP (2013/15115-8), 
EMGDP from  FAPESP (2013/10559-5) and CNPq (306598/2009-4) grants. 
AL acknowledges the support by NASA TCAN 144AAG1967 grant 
and NSF AST 1816234 award.
The Flatiron Institute is supported by the Simons Foundation.
The numerical simulations presented here were performed in the cluster of the Group of Plasmas and High-Energy Astrophysics (GAPAE), acquired with support from FAPESP (grant 2013/10559-5). This work also made use of the computing facilities of the Laboratory of Astroinformatics (IAG/USP, NAT/Unicsul), whose purchase was made possible also by FAPESP (grant 2009/54006-4).

\section*{Data availability}

The data underlying this article will be shared on reasonable request to the corresponding author.








\appendix

\section{The test-field method}
\label{ap:test-field}

The test-field method, developed initially for spherical geodynamo simulations 
\citep{schrinner_etal_2005}, allows to compute unambiguously the contribution of
the small scale on the large scale dynamics. In mean-field analysis this contribution
is described by the electromotive force, 
$\langle {\cal E} \rangle = \langle {\bf u}^{\prime} \times {\bf B}^{\prime} \rangle$,
where the primes denote small-scale fields. Consider that ${\bm u}^{\prime}$ is
the turbulent part of a velocity field, ${\bf u}$, in a given simulation, and also 
a set of test-fields, ${\bm B}^T$ (these fields are independent between themselves 
and of the magnetic field of the simulation ${\bm B}$),  
such that the evolution of ${\bm B}^{\prime}$ may be computed in a set of 
partial differential equations 
which depends only on ${\bf u}$ and ${\bm B}^T$. Thus, it is possible to compute the
electromotive force due to the velocity field as
$\langle {\cal E}^T \rangle = \langle {\bf u}^{\prime} \times {\bf B}^{T \prime} \rangle$.
For mean magnetic field varying slowly in space and time, and systems where 
averages in the $x$ and $y$ directions are meaningful,  it is possible to 
write the electromotive force as 
\begin{equation}
\langle {\cal E}_i \rangle = \alpha_{ij} \langle B_j \rangle - \eta_{ij} \langle J_j \rangle,
\end{equation}
\noindent
with, $i,j = 1,2$. Thus, with the use of 4 test fields it is possible to obtain the
4+4 components of $\alpha$ and $\eta$ \cite[see][]{brandenburg_etal_2008}. 

For the simulations presented in this work, without rotation and with non-helical
unstratified turbulence, the inductive terms in the electromotive force, $\alpha_{ij}$, 
must be zero on average.
Therefore, only 2 test fields would be sufficient to determine the turbulent diffusivity.  
Nevertheless, we use 4 test fields to verify the existence of other turbulent effects.
These are given by
\begin{eqnarray}
\mean{\bm B}^{1c} = B_0(\cos kz, 0,0), \;\;
\mean{\bm B}^{2c} = B_0(0,\cos kz,0),\\\nonumber
\mean{\bm B}^{1s} = B_0(\sin kz,0,0), \; \;
\mean{\bm B}^{2s} = B_0(0, \sin kz,0),\nonumber
\end{eqnarray}
where $k = k_{z, {\rm tf}} = 2 \pi / {\rm L}_{\perp}$. Thus, the $\alpha$-effect is given by the diagonal components 
of $\alpha_{ij}$, $\alpha_{xx} = \alpha_{11}$ and
$\alpha_{yy} = \alpha_{22}$, while the turbulent diffusion is given by 
$\eta_t = \frac{1}{2}(\eta_{11} + \eta_{22})$. In addition, the turbulent driven
advection of the magnetic field, often called turbulent pumping, can be measured by
$\gamma = \frac{1}{2}(\alpha_{21} - \alpha_{12})$. As expected, all the
coefficients but $\eta_t$ are consistent with zero.


\bsp	
\label{lastpage}
\end{document}